\documentclass[pre,floatfix, superscriptaddress,a4paper,nofootinbib]{revtex4}
\usepackage{amsmath,bm,graphicx}
\usepackage{amssymb}
\usepackage{amsfonts}
\usepackage{graphicx}
\usepackage{amssymb}
\usepackage{mathrsfs}
\usepackage{amsmath}
\usepackage{xcolor}
\usepackage{latexsym}
\usepackage{amsfonts}
\usepackage{hyperref}
\usepackage{subfigure}
\usepackage{wasysym}
\usepackage{dcolumn}
\usepackage{bm}
\usepackage{natbib}
\usepackage{soul}
\newcommand{\pa}[1]{\textcolor{black}{#1}}
\usepackage{ulem}

\begin{document}

\title{How Spatially Modulated Activity Reshapes Active Polymer Conformations}

\author{Paolo Malgaretti}
\email[Corresponding Author : ]{p.malgaretti@fz-juelich.de }
\affiliation{ Helmholtz Institute Erlangen-N\"urnberg for Renewable Energy (IET-2),
Forschungszentrum J\"ulich, Cauerstr.~1, 91058 Erlangen, Germany}
\author{Emanuele Locatelli}
\affiliation{Department of Physics and Astronomy, University of Padova, Via Marzolo 8, I-35131 Padova, Italy}
\affiliation{INFN, Sezione di Padova, Via Marzolo 8, I-35131 Padova, Italy}

\begin{abstract}
Active polymers are driven out of equilibrium by internal forces and exhibit conformational properties that differ fundamentally from those of passive chains. Here we study how spatially modulated tangential activity reshapes the conformations of semiflexible polymers. Using a continuum Rouse model with bending rigidity, we develop a systematic expansion in the limit of weak activity and derive analytical expressions for mode correlations, gyration radius, and end-to-end distance under sinusoidally varying propulsion.
We show that spatially structured activity induces a mode-dependent transition between polymer shrinking and swelling. Uniform or low-mode forcing produces compact, globule-like conformations, whereas higher modes generate alternating stretched and compressed segments, leading to globally swollen chains. Different polymer sizes respond differently to activity, allowing for conformations that are compact in gyration radius yet extended in end-to-end distance. Langevin dynamics simulations quantitatively confirm the theoretical predictions. Our results demonstrate that even weak, patterned activity provides a powerful mechanism to control polymer conformations far from equilibrium.
\end{abstract}

\maketitle

\section{Introduction}

Active systems are characterized by their ability to break equilibrium at the local scale. 
This disruption leads to the emergence of dynamical and collective properties that are 
strikingly different from their equilibrium counterparts or even those of externally driven 
systems~\cite{marchetti2013hydrodynamics}. For such reason, they have been the subject of 
intense study for more than two decades. Examples include Motility-Induced Phase Separation 
(MIPS)~\cite{cates2015motility}, activity-driven microphase separation in continuum 
models~\cite{tjhung2018cluster}, giant number fluctuations~\cite{narayan2007long}, active 
turbulence~\cite{wensink2012meso}, self-organisation phenomena~\cite{needleman2017active}, 
and flocking phases~\cite{vicsek1995novel}.

Within active systems, those composed of filamentous units have recently risen in popularity 
because of their relevance in biological systems, spanning different length scales from the 
cellular level to the macro-scale. For example, the cytoskeleton and the intracellular 
trafficking network, powered by molecular motors, serve diverse purposes~\cite{howard2001mechanics, alberts2015essential}. Activity is also relevant in chromatin dynamics at the level of the whole nucleus~\cite{zidovska2013micron, eshghi2023activity}, for interphase organisation~\cite{mahajan2022euchromatin, goychuk2023polymer} with connections to experimental evidence~\cite{carlon2024pre}, and during replication~\cite{narlikar2013mechanisms}. 
At larger scales, cilia arrays, flagella, cyanobacteria and worms are examples of filamentous 
systems that consume energy to perform some form of directed motion~\cite{lauga2009hydrodynamics, faluweki2023active, kurjahn2024collective, rosko2025cellular, Deblais2020, deblais2023worm, sinaasappel2025locomotion}. 
Further, active filaments are relevant beyond biological systems, with recent technological 
progress opening new frontiers for applications, such as artificial active 
flagella~\cite{Dreyfus2005a,chelakkot2014flagellar,vutukuri2017rational,Nishiguchi2018,martinet2025emergent}, chains made of chemically active droplets~\cite{kumar2024emergent, subramaniam2024rigid}, 
and soft robotic systems~\cite{ozkan2021collective, becker2022active}.

Inspired by these diverse examples, researchers have dedicated significant attention to the 
characterisation of simple active polymer models. Typically, an active polymer is a chain of 
``active'' monomers, each capable of generating force and performing autonomous 
motion~\cite{Bianco2018}. In this study, we focus on the case of tangential or 
polar propulsion, where forces are oriented along the local tangent of the polymer 
backbone~\cite{Isele-Holder2015, Bianco2018, Malgaretti2024}. 
From a biophysical perspective, this model mimics the action of molecular motors on 
biofilaments such as actin or microtubules under certain conditions~\cite{howard2001mechanics,Ripoll2025}. 
Moreover, it represents a minimal model for the locomotive mechanism employed by 
filamentous bacteria or worms, where propulsion occurs via segment contraction or lateral 
protrusions~\cite{deblais2023worm}.

It is interesting to note that, in three dimensions, simulations of tangentially active 
polymers show the emergence of a globule-like phase at high activity when the magnitude 
of the active forces is constant along the chain~\cite{Bianco2018}; such a phase was 
not predicted in analytical models that do not fix the force magnitude along the polymer. 
Indeed, this feature makes the model nonlinear and poses a formidable challenge from an 
analytical perspective. In previous work~\cite{Malgaretti2024}, we introduced a set of 
approximations that highlighted the emergence of a more compact phase, in agreement with 
numerical results.

In this work, we \pa{aim at deriving the simplest model capable of reproducing the dependence of the gyration radius, $\mathcal{R}_G^2$, and end-to-end distance, $\mathcal{R}_E^2$, on the active force observed in the numerical simulations~\cite{Isele-Holder2015, Bianco2018}. Accordingly, we  consider }
a generalized 
Rouse model \pa{and} we perform a systematic expansion at small active forces, obtaining analytical 
expressions for quantities such as the gyration radius and the end-to-end distance.
While so far the focus has been on tangential active polymers with active force homogeneously distributed on the backbone, with the possible addition of some noise~\cite{Ripoll2025} or a diblock architecture~\cite{vatin2024conformation, vatin2025upsurge}, here we derive a model where the magnitude of the force can
be characterized by different patterns. In order to better understand the role of the inhomogeneity of the magnitude of the force, in the following we will focus on the case in which the force is distributed in a cosine-shaped way, characterized by a characteristic wavelength.   

We then compare these predictions with numerical simulations for matching cosine-shaped 
activation profiles, which we name ``single mode'' forcing. 
\pa{Such a choice is due to the fact that the conformations 
of the passive case are fully described by basis of cosine-shaped functions; this provides access to clear and interpretable analytical results and enables a direct comparison between the predictions of the model and the numerical data.} 
Even though,  
due to signal-to-noise ratio issues, we could not reach the
small values of the active force required by the expansion, our
numerical results performed for finite values of the active force confirm
the predictions of our model. This shows an interesting robustness of
our mode even at finite values of the active force.
\pa{In the following section (Sec.II) we are going to outline the derivation of such a model. In particular we will report the  main assumptions which highlight the regime of validity of the model. Since the derivation of the model is lengthy, even though rich of physical insight, we resume here the the main steps of this procedure as a guidance for the reader interested in following the derivation as well as to give a glimpse of the underlying reasoning. The reader who is not interested in the derivation may jump directly to the results in Sec.IV whereas those interested in the full derivation can find it in the Appendix.  
The derivation of the model proceed as follows: first, we define our model (see Eq.\eqref{eq:Rouse}), and the Eigenfunction expansion (see Eqs.~\eqref{eq:r_fourier},\eqref{eq:base}), which formally allows to define the steady-state correlation functions $\langle \mathbf{r}_i\cdot \mathbf{r}_j \rangle^\infty$ (see Eq.~\eqref{eq:rirj}) needed to compute the gyration radius (see Eq.~\eqref{eq:def-RG}) and end-to-end distance (see Eq.~\eqref{eq:def-REE_0}). Next, we focus on the case of short persistence lengths which allows for the derivation of closed formulas for the correlation functions (see Eqs.\eqref{eq:exp-text}) which lead to the final expressions for the gyration radius (see Sec.\ref{sec:rg}) and end-to-end distance (see Sec.\ref{sec:re}).}

\section{Continuum Model}
\begin{figure}[h]
    \centering
    \includegraphics[width=0.75\linewidth]{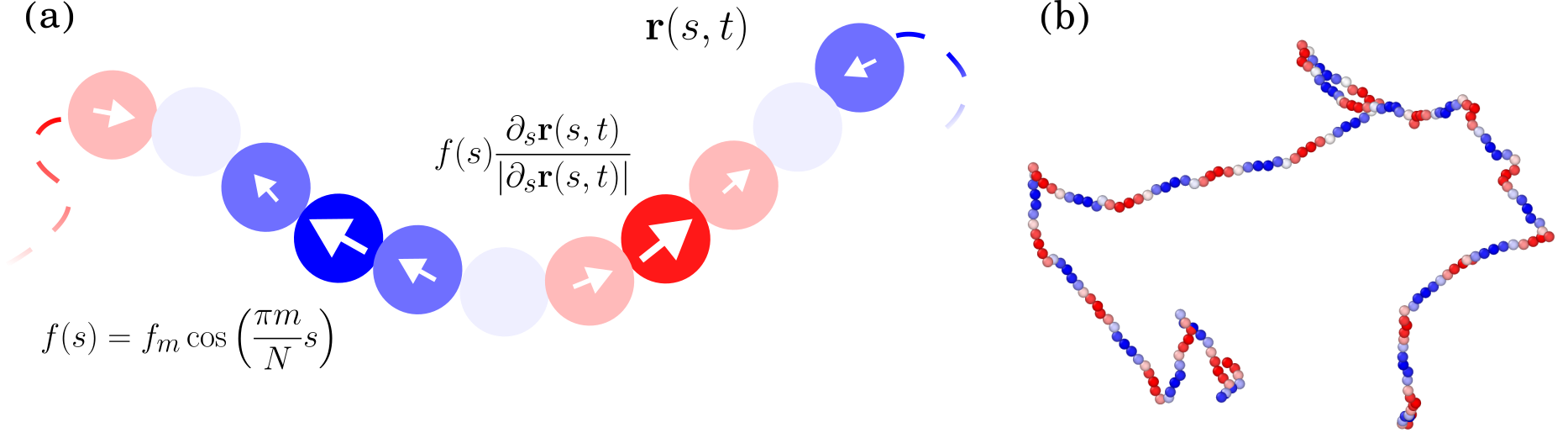}
    \caption{(a) Sketch of the continuum active polymer $\mathbf{r}(s,t)$. The self-propulsion direction is modulated along the backbone as $\cos (\pi ms/N)$ with $s$ the coordinate along the contour of the polymer, $N$ is the degree of polymerization and $m$ an integer; we name this ``single-mode force'' and we highlight the modulation through the verse and size of the arrows. (b) Example snapshot from numerical simulations (see Sec.~\ref{sec:numerical}) with $m=40$; colors and other parameters as in Fig.~\ref{fig:snapshots}.  }
    \label{fig:sketch}
\end{figure}

We model the active polymer via a Rouse model with a finite bending rigidity\pa{, following Ref.~\cite{Malgaretti2024}; the Rouse model is well established in the passive case for where it allows for a simple expansion of the polymer conformations in normal modes.}
The equations of motion read

\begin{equation}
\dot{\mathbf{r}}(s,t)=-\mu G\partial_{s}^{4}\mathbf{r}(s,t)+\mu D\partial_{s}^{2}\mathbf{r}(s,t)+\mu f(s)\frac{\partial_{s}\mathbf{r}(s,t)}{|\partial_{s}\mathbf{r}(s,t)|}+\eta(s,t)\label{eq:Rouse}
\end{equation}
where $s$ is the coordinate along the contour, $\mu$ is the mobility of the monomers and $\boldsymbol{\eta}$ is a white noise
\begin{subequations} \label{eq:model}
\begin{align}
\left\langle \boldsymbol{\eta}(s,t)\right\rangle  & =0\\
\left\langle \eta_{i}(s,t) \eta_{j}(s',t')\right\rangle  & =2\mu k_{B}T\delta_{ij}\delta(t-t')\delta(s-s')
\end{align}
\end{subequations}
delta correlated in time and space. The parameters $D$ and $G$ are related to the monomer size, $b$, and polymer dimensionless persistence length, $l_p$, as 
\begin{equation}
 D=\frac{d}{2}\frac{k_BT}{b^2 l_p}, \qquad G=\frac{d}{2} \frac{k_BT}{b^2} l_p, \qquad l_p=\sqrt{G/D}
 \label{eq:def-D}
\end{equation}
where $d$ is the dimension of the space the polymer is embedded in\pa{; in} this framework, $l_p\in[1:L_0]$, $L_0=N b$ being the contour length of the polymer.
\pa{In} the absence of active force\pa{, that is,} $f(s)=0$, Eq.~\eqref{eq:Rouse} reduces to the standard equilibrium Rouse model with finite bending rigidity~\cite{harnau1995dynamic}.
At variance with Ref.~\cite{Malgaretti2024}, the tangential active force $f(s)$ is not necessarily constant along the contour and is, for the moment, not further \pa{specified. However,} it is still characterized by a typical magnitude, dubbed $f(s)$; see Fig.~\ref{fig:sketch}a) for a sketch of the model.\\
Equation~\eqref{eq:Rouse} should be completed with a set of boundary conditions (applied without the noise~\cite{harnau1995dynamic,Winkler2016}); for the case of free ends, that is, no forces on the head and tail of the polymer, the boundary conditions read~\cite{harnau1995dynamic}:
\begin{align}
 \left[D\partial_s\mathbf{r}(s,t)-G\partial^3_s\mathbf{r}(s,t)\right]_{s=\pm N/2}&=0\label{eq:BC-1}\\
 \left[D_0\partial_s\mathbf{r}(s,t)\pm G\partial^2_s\mathbf{r}(s,t)\right]_{s=\pm N/2}&=0\label{eq:BC-2}
 \end{align}
with $D_0 = Dl_p$. 
Here we focus on the free end case for a two-fold reason: on the one hand, this is the condition that polymers experience at equilibrium and hence it allows us to rely on the previous works done on polymers at equilibrium (for example Ref.~\cite{harnau1995dynamic}). On the other hand, this is the case which has been numerically studied in Ref.~\cite{Bianco2018} where the shrinkage of the gyration radius has been reported. We highlight the fact that the nature of the end beads can strongly influence the structure and dynamics of an active filament, e.g., when pushing a load~\cite{isele2016dynamics,fily2020buckling,maurya2025characteristic} or when the end monomer leads the dynamics~\cite{li2023nonequilibrium, vatin2024conformation,sahoo2025target}.
Notice that Eq.~\eqref{eq:BC-1} is equivalent to request that
\begin{align}
 \dot{\mathbf{r}}_{CM}(t)=\frac{1}{N}\int_{-\frac{N}{2}}^{\frac{N}{2}} \dot{\mathbf{r}}(t)ds=\frac{1}{N}\int_{-\frac{N}{2}}^{\frac{N}{2}} \mu f(s)\frac{\partial_{s}\mathbf{r}(s,t)}{|\partial_{s}\mathbf{r}(s,t)|}ds +\frac{1}{N}\int_{-\frac{N}{2}}^{\frac{N}{2}} \eta(s,t) ds
 \label{eq:r_CM-1}
\end{align}
that is, that the dynamics of the center of mass is governed by the net force on the polymer.
\pa{We remark that the introduction of a small (but finite) bending rigidity is crucial. In fact, as shown in Ref.~\cite{Malgaretti2024}, the dynamics of a flexible Rouse polymer with finite activity strongly depends on the number of Rouse modes that are considered; when infinitely many Rouse modes are accounted for, the denominator in Eq.~\eqref{eq:r_CM-1} diverges and the polymer becomes passive. This is quite inconvenient, since the underlying idea of the Rouse model is to capture the dynamics of the ``real'' polymer by summing over all the modes. In Ref.~\cite{Malgaretti2024}, we showed that  dding a small bending rigidity eventually removes the dependence on the number of modes accounted for, identifying a cutoff mode $n_p=\frac{l_p}{b} = \sqrt{\frac{G}{D b^2}}$ above which  the amplitudes of modes decay as $n^{-4}$, which enforces the convergence of the denominator.}

\subsection{Analytical approach: normal modes, small force, and small persistence length expansion}

\pa{The dynamics of the active polymer is encoded in the stochastic partial differential equation, Eq.~\eqref{eq:Rouse} along with the boundary conditions. However, here we are interested in the average values of  quantities like the gyration radius, $\mathcal{R}_G$, and the module of the end-to-end vector, $\mathcal{R}_E$. For passive Rouse model such averages are obtained by decomposing Eq.~\eqref{eq:Rouse} on the normal modes and then expressing both $\mathcal{R}_G$ and $\mathcal{R}_E$ as a function of the amplitudes of these modes. 
Unfortunately, finding the eigenvectors of the differential operator on the right hand side of Eq.~\eqref{eq:Rouse} is a formidable task due to the non-linearity introduced by the active term. Therefore in the following we will use the eigenvectors of the passive case (see Appendix~\ref{app:eigen}) which, due to the active term, are not orthogonal anymore. At steady state, the gyration radius and the end-to-end distance read (see Sec.\ref{sec:rg} and Sec.\ref{sec:re} for their derivation)
\begin{align}
\mathcal{R}_{G}^2&=\frac{1}{N}\sum_{i\neq 0}{\langle \mathbf{r}_{i}\cdot \mathbf{r}_{i}\rangle^\infty}
\label{eq:def-RG}\\
\mathcal{R}^2_E &= \frac{8}{N}\sum_{i=odd}\sum_{j=odd} \langle \mathbf{r}_{i}\cdot\mathbf{r}_{j}\rangle^\infty\label{eq:def-REE_0}
\end{align}
which are functions of expectation values of the mode correlations at steady state $\langle \mathbf{r}_i\cdot \mathbf{r}_j\rangle^{\infty}$
\begin{align}
    \langle \mathbf{r}_i\cdot \mathbf{r}_j \rangle^\infty=\lim\limits_ {t\rightarrow \infty} \langle \mathbf{r}_i(t)\cdot \mathbf{r}_j(t)\rangle\,.
\end{align}
While the formal expressions of $\langle \mathbf{r}_i\cdot \mathbf{r}_j \rangle^\infty$ can be derived in the general case (see Appendix~\ref{sec:ss_corr}), simple closed  formulas can be derived when both the active force and the persistence length are small. 
The latter leads to decompose both the amplitudes of the "Rouse" modes and the force onto a cosine-base (see Appendxi~\ref{sec:ss_corr}). In order to get further insight, in the following we assume that only one "mode", $f_m$, of the active force is excited (see Appendix~\ref{app:single-mode}). Accordingly, by expanding the model for small values of the active force we identify the relevant dimensionless small parameters as the P\'eclet number defined as
\begin{align}
    \mathrm{Pe}_m = \frac{f_m b\sqrt{N}}{k_BT}
    \label{eq:def-Pe}
\end{align}
where $f_m$ is the amplitude of the "m"th mode of the force. 
The last expression is very interesting since it shows that for active polymers the ``control parameter'', $\mathrm{Pe}_m$ depends on their size, $N$ which has multiple consequences. First, the dependence of $\mathrm{Pe}$ on $m$ does not allow to identify scaling behavior upon increasing $N$ at given  $\mathrm{Pe}$. Second, from a thermodynamic perspective, $N$ is an ``extensive'' quantity. This implies that the dimensionless parameter governing the onset of higher order contributions depends not only on ``intensive'' quantities such as the monomer length, $b$, and the strength of the force, $f_m$, but also on their size.
It is interesting to note that such a P\'eclet number has been also empirically derived in Ref.~\cite{Malgaretti2024}.  
While the full derivation is in the Appendix, here we discuss some features of the amplitudes of the correlation functions which will be important for $\mathcal{R}_G$ and $\mathcal{R}_E$.  
\begin{subequations}\label{eq:exp-text}
\begin{align}
    \langle \mathbf{r}_i\cdot \mathbf{r}_j\rangle^\infty_0 =& \delta_{ij}d\frac{k_BT N^2}{\pi^2 D}\frac{1}{\bar{z}_i}\label{eq:rr0-final}\\
    \langle \mathbf{r}_i\cdot \mathbf{r}_j\rangle^\infty_1 =&-8\sqrt{2}\frac{b^2 N^2}{\pi^4 d^2}\frac{1}{l_p^2} \frac{(1-\delta_{ij})}{\bar{z}_i+\bar{z}_j} \ \mathrm{Pe}_m\left(\frac{\bar{\xi}_{ijm}^{(0)}}{\bar{z}_j}+\frac{\bar{\xi}_{jim}^{(0)}}{\bar{z}_i}\right)\label{eq:rr1-final}\\
    \left\langle \mathbf{r}_i\cdot \mathbf{r}_j\right\rangle^\infty_2 =& \frac{64}{\pi^6}\frac{b^2 N^2}{l_p d^3}\frac{1}{\bar{z}_i+\bar{z}_j} \mathrm{Pe}^2_{m} \left[\sum_{n\neq i} \frac{1-\delta_{jn}}{\bar{z}_n+\bar{z}_j}\left(\frac{\bar{\xi}^{(0)}_{jnm}}{\bar{z}_n}+\frac{\bar{\xi}^{(0)}_{njm}}{\bar{z}_j}\right) \bar{\xi}^{(0)}_{inm}+\sum_{n\neq j} \frac{1-\delta_{in}}{\bar{z}_n+\bar{z}_i}\left(\frac{\bar{\xi}^{(0)}_{inm}}{\bar{z}_n}+\frac{\bar{\xi}^{(0)}_{nim}}{\bar{z}_i}\right) \bar{\xi}^{(0)}_{jnm}\right]+\nonumber\\
&-\frac{8 b^2 N^2}{d \pi^6}\frac{1-\delta_{ij}}{\bar{z}_i+\bar{z}_j} \mathrm{Pe}_m \left[\frac{\bar{\xi}^{(1)}_{ijm}}{\bar{z}_j}+\frac{\bar{\xi}^{(1)}_{jim}}{\bar{z}_i}\right]-\delta_{ij}\frac{8 b^2 N^2}{\pi^6 d}\frac{1}{\bar{z}_i^2}  \mathrm{Pe}_m \bar{\xi}^{(1)}_{iim}\label{eq:rr2-final}
\end{align}
\end{subequations}
where 
\begin{align}
    \bar{z}_i = i^2(1+\Lambda_P i^2)
\end{align}
are the normalized eigenvalues and $\xi^{(0)},\xi^{(1)}$ are some trigonometric integrals defined in Appendix~\ref{app:1}. Eqs.~\eqref{eq:exp-text} shows two main interesting features. First, at zeroth order, Eq.~\eqref{eq:rr0-final} retrieves the standard amplitudes of the Rouse modes for a ghost polymers with vanishing bending rigidity. We also remark that in the passive case the modes are orthogonal to each other. Second, at first order, the diagonal terms $\left\langle \mathbf{r}_i\cdot \mathbf{r}_j\right\rangle^\infty_1$ of the correlation matrix are all zero, as shown in Fig.~\ref{fig:rr1}. Moreover, while this property is independent of the force mode, $m$, upon increasing $m$ the region of null correlation about the main diagonal widens.} 

\section{Numerical model and simulation details}
\label{sec:numerical}
We model active polymer chains as semi-flexible Gaussian beads-spring linear chains consisting of $N$ monomers, suspended in a 3D bulk fluid. Since we consider Gaussian polymers, non-neighboring monomers do not interact with each other. To guarantee chain connectivity, neighboring monomers along the chain interact via the combination of the truncated and shifted Lennard-Jones potential (also known as WCA) and the Finitely Extensible Nonlinear Elastic (FENE) potential. The truncated and shifted Lennard-Jones (LJ) potential reads
\begin{equation}
V_{\text{LJ}}(r) =
\begin{cases}
  4\epsilon \left[ \left(\frac{\sigma}{r} \right)^{12}- \left(\frac{\sigma}{r}\right)^{6}+\frac{1}{4}\right] & \text{for~}  r < 2^{1/6}\sigma \\
  0 & \text{for~} r \geq 2^{1/6}\sigma
\end{cases}
\end{equation}
where $\sigma=1$ is the diameter of the monomer and is taken as the unit of length, $\epsilon=10\,k_B T$ with $k_B T=1$ the energy scale and $r=|\vec{r}_i - \vec{r}_j|$ is the Euclidean distance between the monomers $i$ and $j$ positioned at $\vec{r}_i$ and $\vec{r}_j$, respectively. The Finitely Extensible Nonlinear Elastic (FENE) potential reads
\begin{equation}
	V_\mathrm{FENE}(r) = -\frac{K r_0^2}{2} \ln \left[ 1 - \left( \frac{r}{r_\mathrm{0}} \right)^2 \right]
\end{equation}
with $K= 30\,\epsilon/\sigma^2=300\,k_\mathrm{B}T/\sigma^2$ and $r_0=1.5\sigma$. The combination of both FENE and WCA is necessary for neighboring monomers along the chain, as it provides a typical bond length $b \simeq 0.97 \sigma$ and, with the chosen parameters, the bond length remains fixed even at the highest values of the active force considered in this work. 
The polymer rigidity is introduced with the bending potential:
\begin{equation}
\label{Ub}
U_{b} (\theta) = \kappa (1+\cos{\theta}),
\end{equation}
where $\theta$ is the angle formed by three consecutive beads along the backbone and $\kappa=3 k_B T$ is the bending energy; this results in a  persistence length ${l_p}/{L} = {\kappa }/{(k_B T N)} = 0.0075$, for $N=400$.\\
Activity is introduced as a tangential self-propulsion\cite{Bianco2018}, where an active monomer $i$ at position $\vec{r}_{i}$ is subject to an active force $\vec{f}_i$ parallel to the normalized tangent vector $\hat{t}_i = (\vec{r}_{i+1} - \vec{r}_{i-1})/|\vec{r}_{i+1} - \vec{r}_{i-1}|$; the end monomers are always passive. In this work, we modulate the force on each bead via a sinusoidal function of the bead index along the polymer, that can be tuned by a parameter $m$, assuming only integer values ($m \in \mathbb{N}$). Overall, the active force on bead $2 \leq i \leq N-1$ reads 
\pa{
\begin{equation}
\vec{f}_{m} (i) = f_m \cos \left( k_m i \right) \hat{t}_i.  
\label{eq:single_mode}
\end{equation}
where $f_m$ is the strength of the activity} \pa{and $k_m = \frac{\pi m}{N}$}. \pa{Equation~\eqref{eq:single_mode} is, thus, Eq.~\eqref{eq:f_single} for a bead-spring chain.} 
Notice that with this prescription, the magnitude of the active force is not constant along the chain for $m \neq 0$.\\
We simulate polymers in bulk conditions via periodic boundary conditions, employing the open source code LAMMPS\cite{thompson2022lammps}, with in-house modifications to implement the tangential activity Eq.~\eqref{eq:single_mode}. Langevin Dynamics simulations are performed, disregarding hydrodynamic interactions. The equations of motion are integrated using the velocity Verlet algorithm, with an elementary time step $\Delta t = 10^{-3}$. Units of mass, length, and energy are set to $\overline{m}=1$, $\sigma=1$, and $k_B T = 1$, respectively. The unit of time is $\tau= \sqrt{\overline{m} \sigma^2/ k_B T} = 1$. The overdamped regime is ensured by setting the friction coefficient $\gamma=20 \, k_B T \, \tau/\sigma^2$ for $f_m>1$ \cite{fazelzadeh2022effects}; otherwise, we set $\gamma=1 \, k_B T \, \tau/\sigma^2$.
We perform simulations varying the activity $f_m = 0.001, 0.01, 1, 10 \,k_BT/\sigma$, the polymer length $N=200,400$ and the force mode $m=0,1,2,3,4,5,6,7,8,11,14,20,40,100$; we collect data over $M=25$ independent trajectories for each set of parameters. After reaching a steady state, production runs are performed for 
$\simeq 2 \cdot 10^8$ time steps ($2\cdot 10^5 \tau$) and the polymer conformations are sampled at a rate of $10^6$ time steps ($2\cdot 10^3 \tau$). 

\begin{figure}
    \centering
    \includegraphics[width=0.9\textwidth]{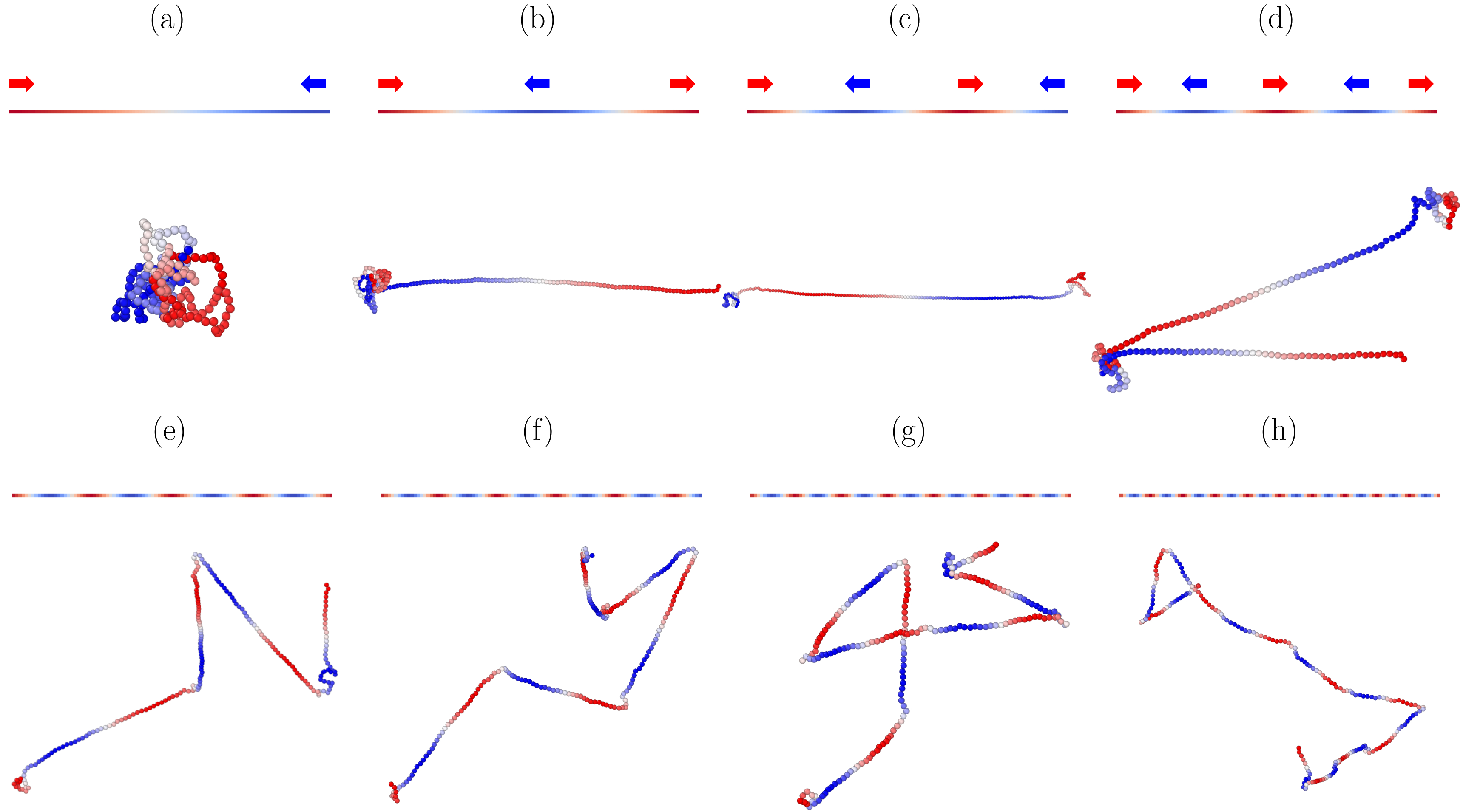}
    \caption{Snapshots of active polymers, $N=200$, $\mathrm{Pe}_m=10$ and different values of the \pa{force mode} $m$. The chosen color map highlights both the sign and magnitude of the active force, shades of red being positive, shades of blue being negative and white being zero; arrows indicate the overall direction of the force in the section of the chain close to the arrow. (a) $m=1$, (b) $m=2$, (c) $m=3$, (d) $m=4$, (e) $m=8$, (f) $m=11$, (g) $m=14$, (h) $m=20$.}
    \label{fig:snapshots}
\end{figure}

\section{Results}

%
Before carrying on a direct comparison between theory and simulation results, we briefly show and discuss the phenomenology that emerges from simulations. Figure~\ref{fig:snapshots} as well as Fig.~\ref{fig:sketch}b) report snapshot of polymer conformations from numerical simulations of fixed degree of polymerization $N=200$ beads and $\mathrm{Pe}_m=141.42$. We remark that our theoretical approach is valid at sufficiently low activity; nevertheless, the phenomenology is more evident at high values of $\mathrm{Pe}_m$ and, for such reason, we use this case for showcasing the effect of the active force modulation. The snapshots are color-coded to highlight the force modulation: the color code highlights both the sign and magnitude of the active force, shades of red being positive, shades of blue being negative and white being zero. The force modulation is then visible on the conformations and is explicitly reported on the colorbars above the snapshots. At low values of $m$ we add arrows above the colorbars to indicate the direction of the active force (see Fig.~\ref{fig:snapshots}); for large values of $m$, this representation becomes impractical and, thus, arrows are omitted \pa{in the bottom row of Fig.~\ref{fig:snapshots}}.

First, we notice that, in the case $m=1$ (Fig.~\ref{fig:snapshots}a), the polymer conformation is considerably more compact than in all the other cases and reminds of the globule-like conformations observed in a tangentially active polymer~\cite{Bianco2018}, which is recovered in this framework as the case $m=0$. However, for $m>1$ the conformations appear much more elongated, with visibly stretched and visibly compact regions, especially at low values of $m$. The snapshots suggests that the appearance of these regions depends on the alternation of the direction of propulsion, that is, on the alternation of positive and negative propulsion sections. If, proceeding along the chain, a negative section comes after a positive one, the corresponding polymer region is collapsed, as the active forces (showed with arrows in Fig. ~\ref{fig:snapshots}) point towards each other, causing the filament to buckle. On the contrary, if a positive section follows a negative one the corresponding polymer region is stretched as the active forces point away from each other. The higher the value of $m$, the more stretched sections the polymer can develop; since the polymer is Gaussian, the stretched regions are independent from each other and can arrange freely in space (see Fig.~\ref{fig:sketch}b and \ref{fig:snapshots}).  
In the following we analyze the dependence of the gyration radius and the end-to-end \pa{distance} on the distribution of the active forces. While for passive polymers these observables have a very similar dependence on the microscopic parameters characterizing the polymer \pa{our results suggest that for active polymers it is not so}. \pa{Interestingly, we do not impose a block structure of the active force, which could justify the breakdown of the relationship between these two quantities, but rather a continuous, oscillating profile.}  \pa{For these reasons} we analyze them separately.

\subsection{Gyration Radius}
\label{sec:rg}
We now turn to the typical quantities that characterize the conformations of the polymers. 
\pa{As derived in Appendix~\ref{app:der-RG} at steady state the gyration radius  reads
\begin{align}
\mathcal{R}^2_G= \frac{1}{N}\sum_{i\neq 0}\langle \mathbf{r}_{i}\cdot \mathbf{r}_{i}\rangle^\infty
\end{align}}
At zeroth order \pa{in the external force (see the expansion in Sec.~\ref{sec:expansion_lp}),} we have
\begin{align}
 \mathcal{R}^2_{G,0}&=\pa{\frac{1}{N}\sum_{i\neq 0}{\langle \mathbf{r}_{i}\cdot \mathbf{r}_{i}\rangle^\infty_0}} = \frac{d}{6}\frac{k_BT N}{D \pi^2} \left(3 \Lambda_p-3 \pi \Lambda_p \coth \left(\frac{\pi }{\Lambda_p}\right)+\pi ^2\right)\cr &=\frac{2l_p b^2 N}{6\pi^2}\left(3 \Lambda_p-3 \pi \Lambda_p \coth \left(\frac{\pi }{\Lambda_p}\right)+\pi ^2\right)
  \label{eq:RG_0}
\end{align}
where we have used Eq.~\eqref{eq:def-D}; note that for $\Lambda_p\rightarrow 0$ (that is, for $N\to \infty$\pa{, see Eq.~\eqref{eq:def-Lambda}}) 
we retrieve $\mathcal{R}^2_{G,0} \simeq \frac{1}{6} b^2 N$, that is, the zeroth order of our expansion correctly yields the equilibrium value at $f_m=0$.
\begin{figure}[t]
    \centering
    \includegraphics[width=0.32\textwidth]{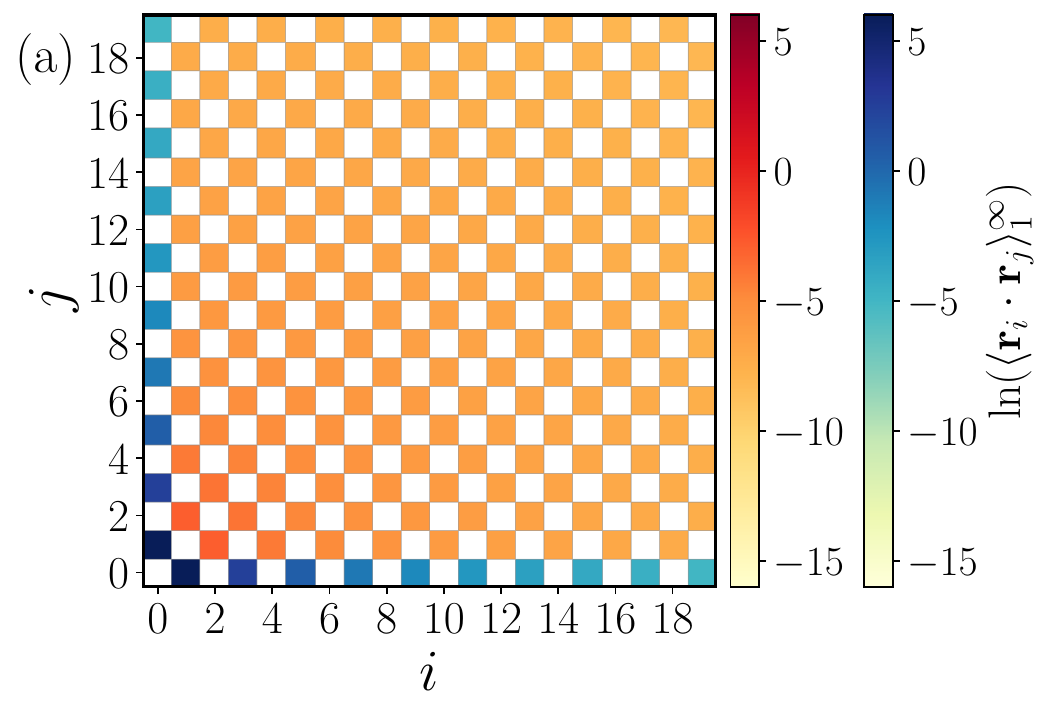}
    \includegraphics[width=0.32\textwidth]{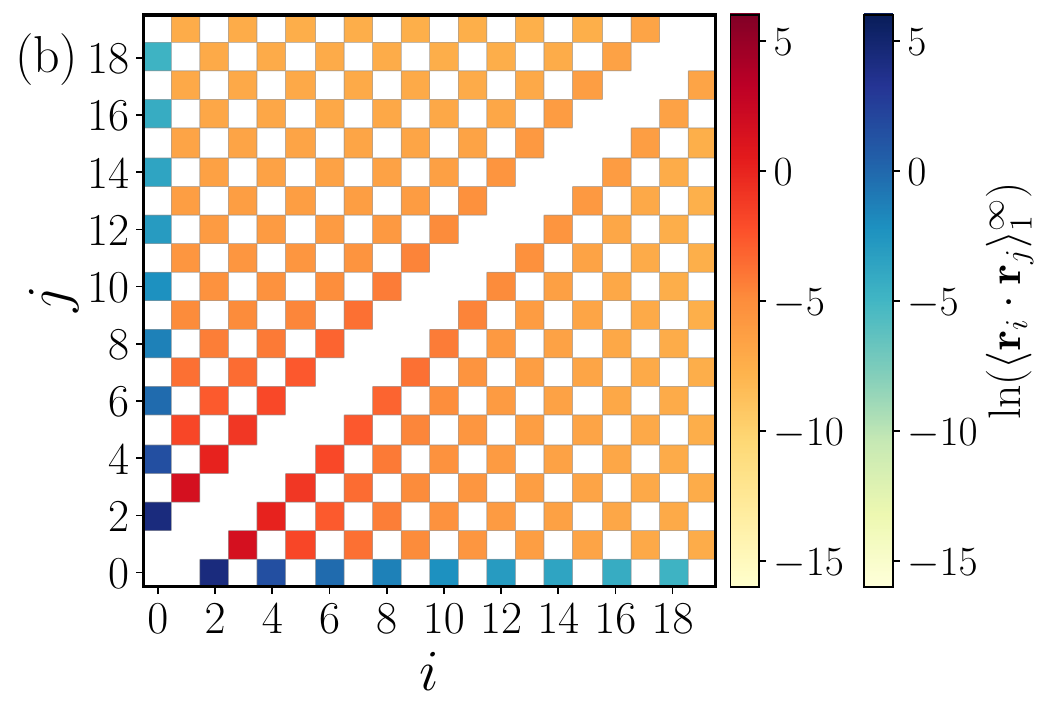}
    \includegraphics[width=0.32\textwidth]{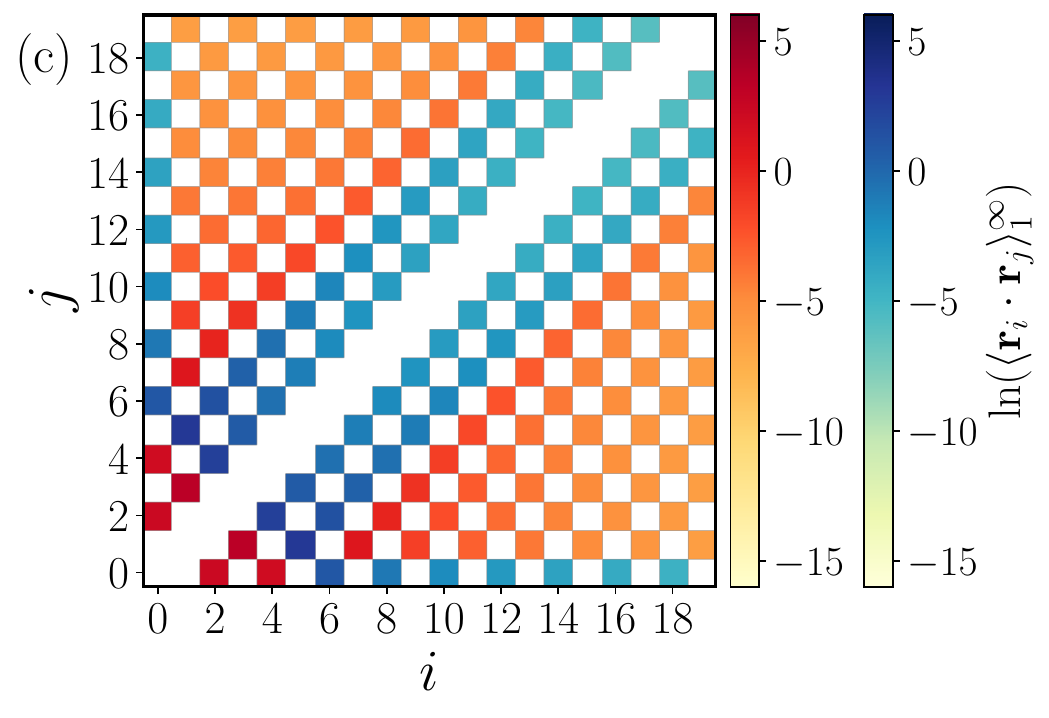}
    \caption{\pa{Mode correlations at first order in the expansion} $\langle \mathbf{r}_i\cdot \mathbf{r}_j\rangle^\infty_1$ \pa{(see Eq.~\eqref{eq:rr1-final}} with \pa{dimensionless persistence length} $l_p = 0.5$, $N=100$, $f_m = 1$ and \pa{force mode} (a) $m=0$, (b) $m=1$, (c) $m=5$. The color codes represents the logarithm of the magnitude of the correlations, using warm colors for positive magnitudes and cool colors for negative magnitudes.}
    \label{fig:rr1}
\end{figure}
\begin{figure}[h]
    \centering
    \includegraphics[width=0.32\textwidth]{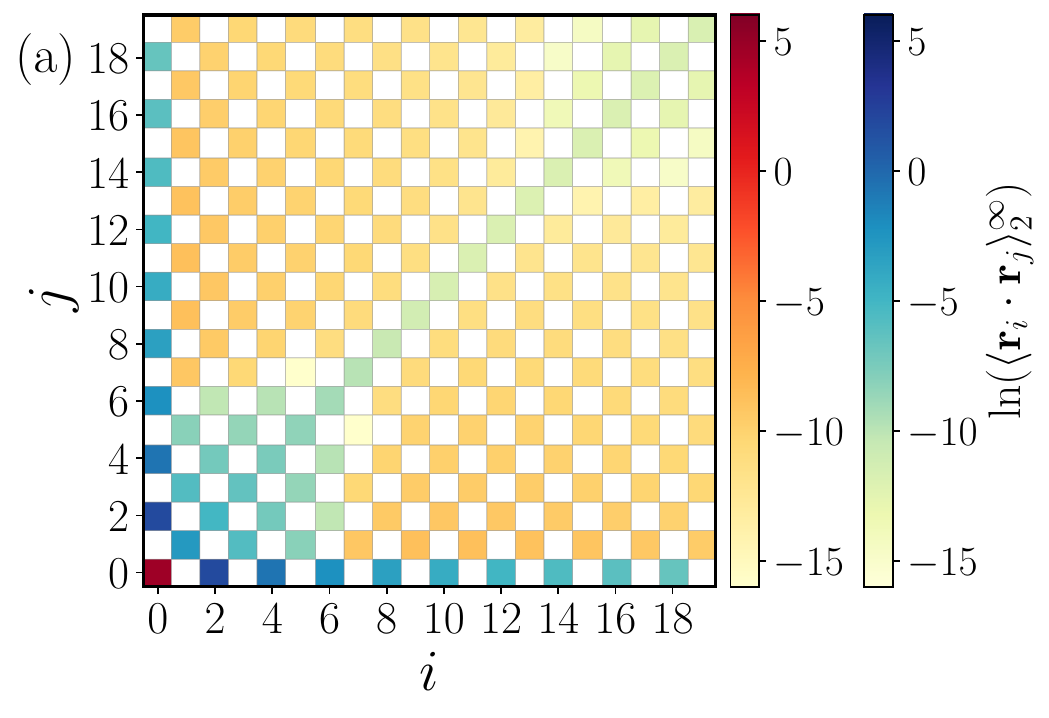}
    \includegraphics[width=0.32\textwidth]{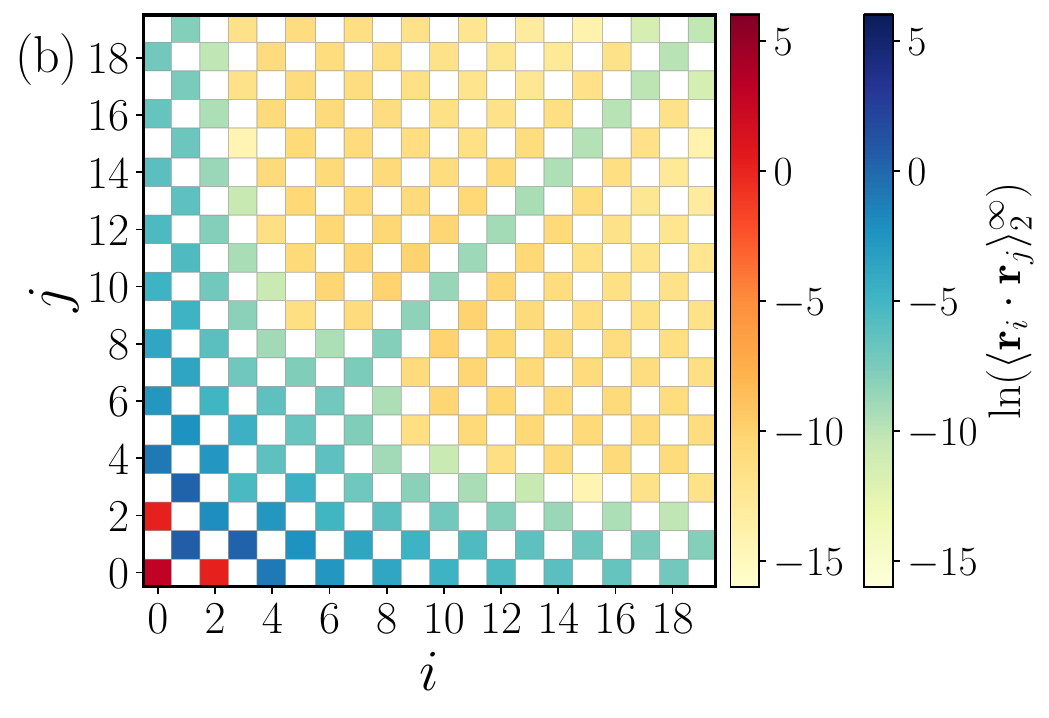}
    \includegraphics[width=0.32\textwidth]{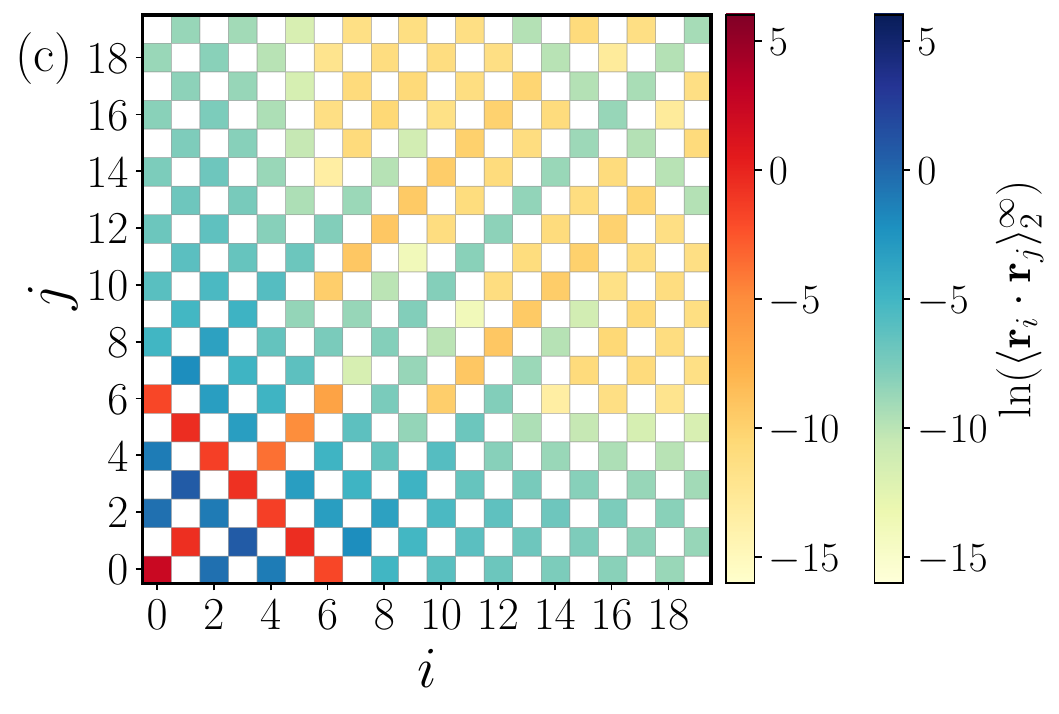}
    \caption{\pa{Mode correlations at second order in the expansion} $\langle \mathbf{r}_i\cdot \mathbf{r}_j\rangle^\infty_2$ \pa{(see Eq.~\eqref{eq:rr2-final}} with \pa{dimensionless persistence length} $l_p = 0.5$, $N=100$, $f_m = 1$ and \pa{force mode} (a) $m=0$, (b) $m=1$, (c) $m=5$. The color codes represents the logarithm of the magnitude of the correlations, using warm colors for positive magnitudes and cool colors for negative magnitudes. }
    \label{fig:rr2}
\end{figure}
At first order, the diagonal 
\pa{correlations} are null, \pa{that is}, $\langle \mathbf{r}_{i}\cdot\mathbf{r}_{i} \rangle_1^\infty=0$ (see  
Eq.~\eqref{eq:o_1-1}  and Fig.\ref{fig:rr1}), and 
\pa{\begin{align}
\mathcal{R}^2_{G,1}=\frac{1}{N}\sum{\langle \mathbf{r}_{i}\cdot\mathbf{r}_{i}\rangle^\infty_1}=0.
\end{align} }
At second order, the correction to $\mathcal{R}^2_G$ depends on the extensibility of the polymer. As we discuss in Appendix~\ref{sec:expansion_Ls}, in order to ensure that the polymer length does not vary upon applying the active force, the stretching coefficient $D$ has to be adjusted (see Eq.~\eqref{eq:def-exp_D}). Accordingly, by substituting Eq.~\eqref{eq:def-exp_D} into Eq.~\eqref{eq:RG_0} we get\footnote{{In principle the same rationale applies also at first order. However, $D_1=0$ and there is no contribution from the corrections to the stretching coefficient at first order.}}:
\begin{align}
 \mathcal{R}^2_{G,D}=-\frac{\overline{D}_2}{D}\mathcal{R}^2_{G,0}=-\frac{1}{6}\frac{d k_BT N}{D}\frac{\overline{D}_2}{D}
 \label{eq:RG_D}
\end{align}
where in the last step we took the limit $\Lambda_p \ll 1$.
Accordingly, at second order, we get
\begin{align}
 \mathcal{R}^2_{G,2}&=\frac{1}{2N}\sum_{i} \langle \mathbf{r}_{i}\cdot \mathbf{r}_{i}\rangle_2^\infty  -\frac{1}{6}\frac{d k_BT N}{D}\frac{\overline{D}_2}{D}
 \label{eq:RG2-0}
\end{align}
\begin{figure}
 \includegraphics[width=0.32\textwidth]{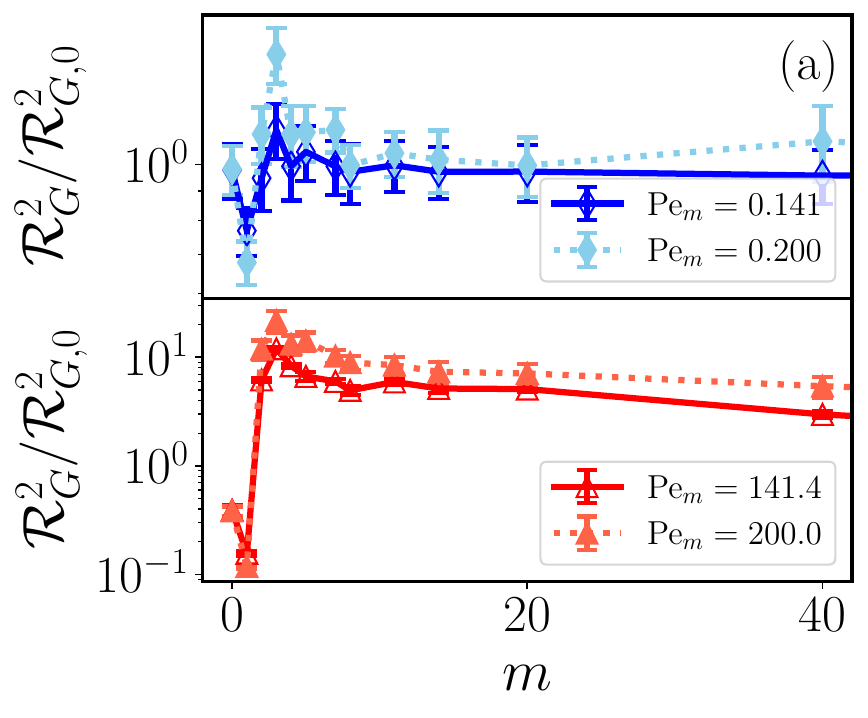}
 \includegraphics[width=0.32\textwidth]{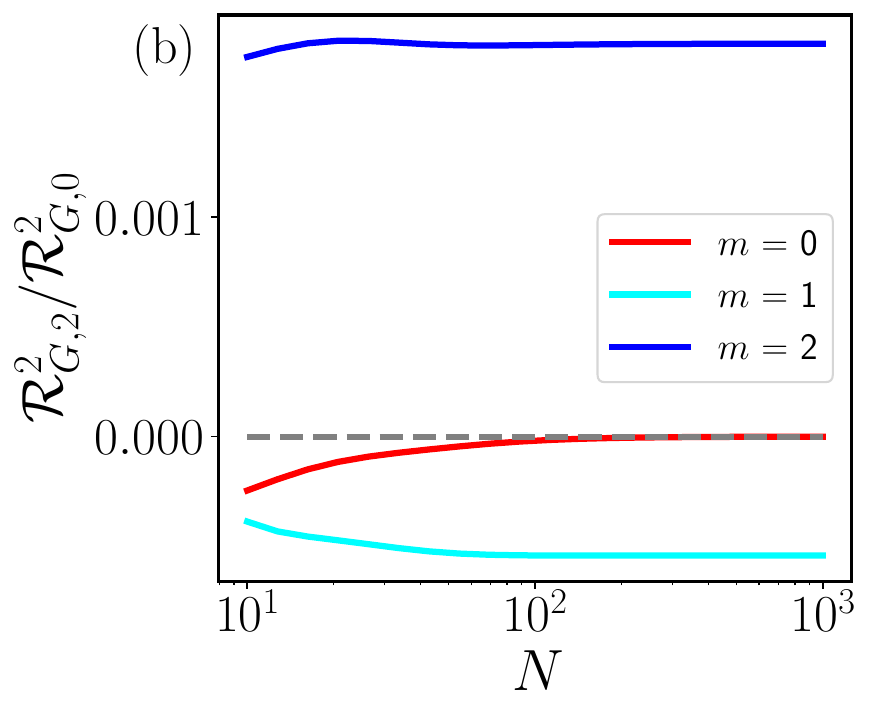}
 \includegraphics[width=0.32\textwidth]{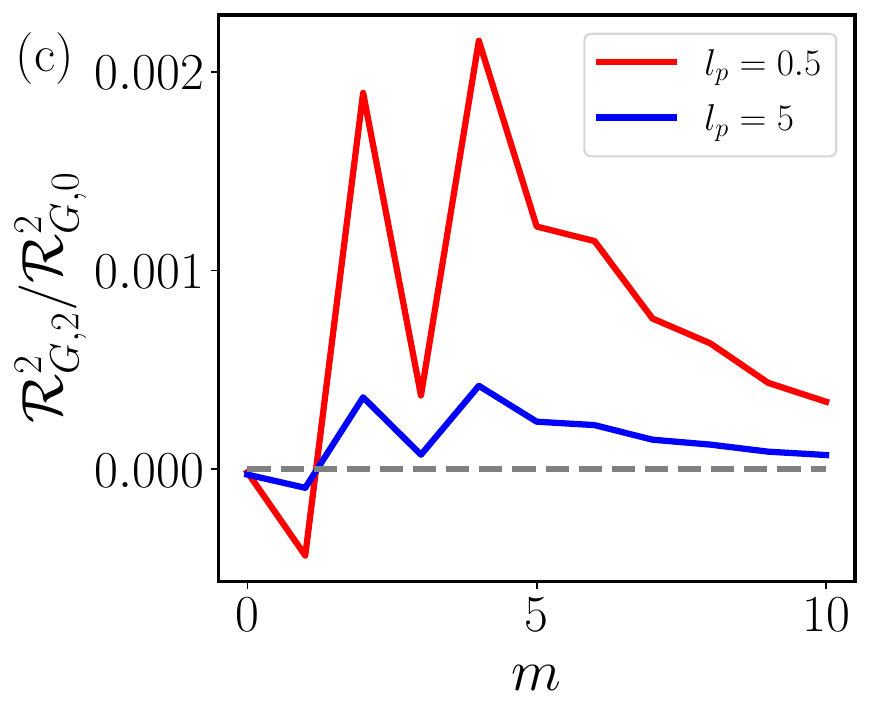}
 \caption{(a) Ratio of the \pa{squared} gyration radius, over the \pa{passive}, equilibrium value $\mathcal{R}^2_{G}/\mathcal{R}^2_{G,0}$ as a function of the \pa{force mode} $m$ and different values of $\mathrm{Pe}_m$ (with full lines $N=200$ and dotted lines $N=400$) from numerical simulations. (b) \pa{Second-order contribution to the squared gyration radius} $\mathcal{R}^2_{G,2}$ (Eq.~\eqref{eq:RG2-0}), normalized by $\mathcal{R}^2_{G,0}$, as a function of $N$ for and $l_p=0.5$, \pa{$\mathrm{Pe}_m=1$} and various values of the force mode $m$; (c) $\mathcal{R}^2_{G,2}/\mathcal{R}^2_{G,0}$ as function of the \pa{force} mode $m$, for $N=100$, \pa{$\mathrm{Pe}_m=1$} and different values of $l_p$.}
 \label{fig:RG}
\end{figure}

The magnitudes of $\langle \mathbf{r}_i \cdot \mathbf{r}_j \rangle_2^\infty$ are shown in Fig.\ref{fig:rr2}. At variance to first order corrections, at second order the diagonal terms are not vanishing and hence they are responsible to the leading corrections to the gyration radius. Notice that, for some combinations of the ``polymer'' mode $i$ and of the ``force'' mode $m$, $\langle \mathbf{r}_i \cdot \mathbf{r}_i \rangle_2^\infty$ attains large negative values (color coded in red) whereas the majority of $i$ and $m$ combinations leads to positive values (color coded in green-yellow). This already suggests that, for some values of $m$, the theoretical expansion could predict a reduction of $\mathcal{R}_G$.

We obtain the prediction for $\mathcal{R}_G^2$ using the numerical values shown in Fig.~\ref{fig:rr2}. Fig.~\ref{fig:RG}a reports the ratio of the squared gyration radius $\mathcal{R}^2_G$ over its equilibrium counterpart $\mathcal{R}^2_{G,0}$,  at fixed active force ($f_m=0.01, 10\, k_BT/\sigma$) for polymers of length $N=200,400$ beads, so that the value of $\mathrm{Pe}_m$ varies for the two different polymers. In both the upper and lower sub-panels, we find a phenomenology consistent with our previous observations: for $m\leq1$ the polymer shrinks ($\mathcal{R}^2_G/\mathcal{R}^2_{G,0} <1$) and for $m>1$ the polymer swells ($\mathcal{R}^2_G/\mathcal{R}^2_{G,0} >1$). Notice that,  for large values of $m$ $\mathcal{R}^2_G/\mathcal{R}^2_{G,0} \simeq 1$: in this limit the polymer becomes a succession of short stretched regions, independent from each other, that can arrange freely in space, producing a coil-like conformation.

In order to compare numerical and theoretical results, one has to evaluate the corrections to the zeroth order, that is Eq.~\eqref{eq:RG2-0}. We show the relative importance of this second-order correction, with respect to the zero-order one, for $\mathrm{Pe}_m = 1$ in Fig.~\ref{fig:RG}b,c as a function of $N$ and $m$, respectively. Interestingly, at fixed $\mathrm{Pe}_m$, the correction becomes constant or negligible with increasing $N$; in simulations, as we do not change the active force to keep $\mathrm{Pe}_m$ fixed, we see the effects of the activity are more pronounced for larger polymers, in agreement with the theoretical trends. Further, notice that the corrections for $m=0$ are negative, that is, the gyration radius decreases when $f_m \neq 0$, in qualitative agreement with previous numerical results~\cite{Bianco2018}. For $m=1$ the correction remains negative, meaning that the model predicts a shrinking of $\mathcal{R}^2_G$, as reported in simulations. Instead, with increasing $m$, the correction becomes positive, that is, the gyration radius increases as observed, again, in simulations. Furthermore, at large values of $m$, the correction becomes negligible, again in qualitative agreement with simulations. We highlight that the theory predicts a larger increase of $\mathcal{R}^2_G$ for $m=2,4$, whereas simulations data report a maximum of $\mathcal{R}^2_G$ for $m=3$; this discrepancy could be caused by the use, in the numerical model, of anharmonic springs (FENE) or by the impossibility to include the correct boundary conditions in the theoretical model.

The fact that the leading corrections are quadratic in the amplitude of the active force implies that the sign of the active force is not relevant. For $m=0$ switching the sign of the active force just means reverting the ``head'' and the ``tail'' of the polymer; as such, it cannot generate any conformational change. In contrast, for $m\neq 0$, and in particular for odd values of $m$, this sign flip \textit{a priori} may be important. For example, for $m=1$, a positive value of the amplitude of the force, $f_1 > 0$, implies that both ends of the polymer experience a force which is pointing ``inwards'' (that is, an overall compression), indeed leading to compact conformations, as in
Fig.~\ref{fig:snapshots}a. However, switching to $f_1<0$ leads to forces pointing outwards. While in the numerical simulations performed at finite P\'eclet we do observe a difference in the gyration radius upon flipping the sign of $f_1$ (see Appendix~\ref{sec:appendix_neg}), this is not the case for the analytical model, which is derived for very weak active forces.

\subsection{End-to-end \pa{distance}}
\label{sec:re}
The second and last conformational quantities that we consider is the square end-to-end distance at steady state (see Appendix~\ref{app:der-RG} for its derivation):
\begin{align}
\mathcal{R}^2_E  =  \frac{8}{N}\sum_{i=odd}\sum_{j=odd} \langle \mathbf{r}_{i}\cdot\mathbf{r}_{j}\rangle^\infty
 \label{eq:def-REE}
\end{align}
At zeroth order\pa{, using Eq.~\eqref{eq:rr0-final},} we have:
\begin{align}
 \mathcal{R}^2_{E,0} &=\frac{8}{N}\sum_{i=odd}\sum_{j=odd} \langle \mathbf{r}_{i}\cdot\mathbf{r}_{j}\rangle^\infty_0
  = \frac{d N k_BT}{\pi D} \left(\pi- 2 \Lambda_p \tanh\left(\frac{\pi}{2\Lambda_p}\right)\right)
 \label{eq:RE_0}
\end{align}
where we use the definitions of $Q$ and $\Lambda_p$, Eqs.~\eqref{eq:def-Lambda}; note that for $\Lambda_p\rightarrow 0$ and $d=3$ we retrieve the Rouse scaling $\mathcal{R}^2_{E,0} \simeq b^2 N$ and the usual relation $\mathcal{R}^2_{E,0}=6\mathcal{R}^2_{G,0}$.

\begin{figure}
    \centering
    \includegraphics[width=0.3\textwidth]{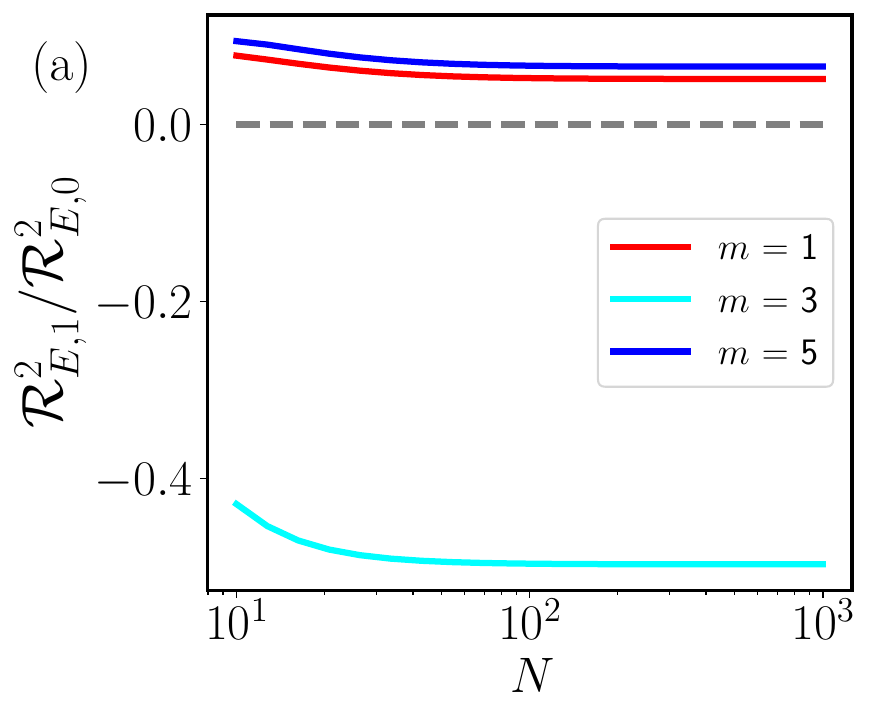}
    \includegraphics[width=0.3\textwidth]{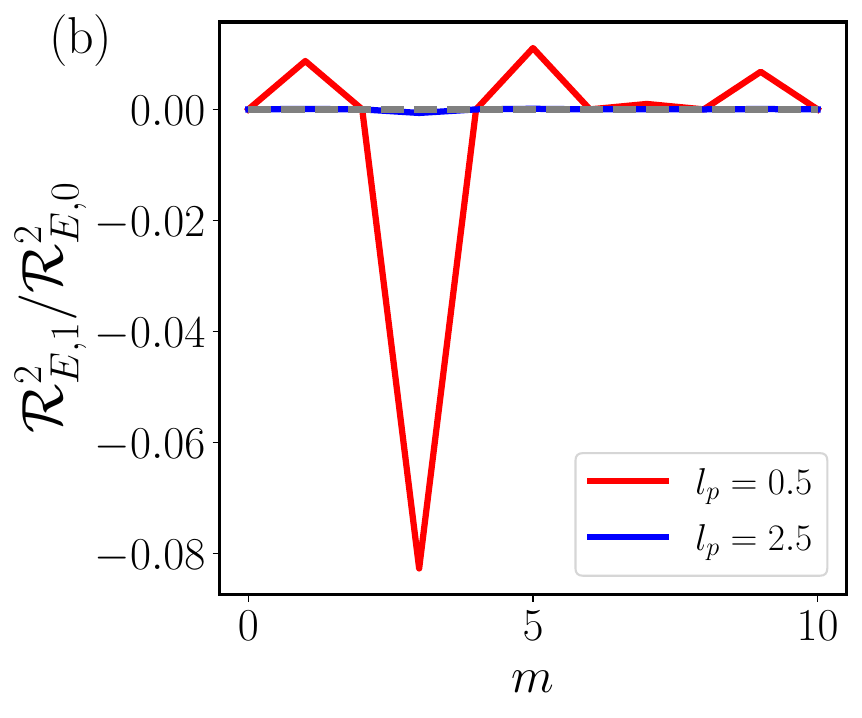}
    \caption{\pa{First-order contribution to the end-to-end square distance} $\mathcal{R}^2_{E,1}$ (Eq.~\eqref{eq:RE_1}), normalized by the \pa{passive} equilibrium end-to-end square distance, $\mathcal{R}^2_{E,0}$, for a single-mode force with $\mathrm{Pe}_m = 1$ (a) $\mathcal{R}^2_{E,1}/\mathcal{R}^2_{E,0}$ as function of $N$ for a single-mode force with $l_p=0.5$ and different values of \pa{the force mode} $m$; (b) $\mathcal{R}^2_{E,1}/\mathcal{R}^2_{E,0}$ as function of $m$ for a single-mode force with $N=100$ and different values of $l_p$.}
    \label{fig:RE1}
\end{figure}

We highlight that the first-order contribution depends on the parity of the active force. For active forces composed solely of the \textit{even} modes, from Eq.~\eqref{eq:rr1} we have $\mathcal{R}^2_{E,1}=0$. In contrast, for active forces that accounts also for \textit{odd} modes we have
\begin{align}
 \mathcal{R}^2_{E,1}=\frac{8}{N}\sum_{i=odd}\sum_{j=odd}\langle \mathbf{r}_{i}\cdot\mathbf{r}_{j}\rangle^\infty_1
\label{eq:RE_1}
\end{align}
We remark that, as shown in Fig.\ref{fig:rr1}b, in the case of odd modes, $m$, of the active force we have $\langle \mathbf{r}_{i}\cdot\mathbf{r}_{j}\rangle^\infty_1 \neq 0$ when both $i$, and $j$ are odds. 
In particular, the sign of $\mathcal{R}^2_{E,1}$ changes\footnote{\pa{We recall that $\mathcal{R}^2_{E,1}$ is the first order correction to $\mathcal{R}^2_E$; hence it can be positive or negative}} upon flipping the sign of $f_m$, with $m=odd$, in Eq.~\eqref{eq:f_exp}; results are reported in Fig.~\ref{fig:RE1}. 
Surprinsingly, $\mathcal{R}^2_{E,1}<0$ for $m=3$, whereas $\mathcal{R}^2_{E,1}>0$ for $m=1$ and $m\geq 5$. 
This results is quite interesting since it highlights a structural difference between the end-to-end \pa{distance} and the gyration radius. In fact, while the former may have leading corrections which are linear in the amplitude of the force, and hence are sensitive to the sign, the leading corrections to the latter are quadratic in the force and hence insensitive to a sign flip of $f_m$.%

For active forces that are composed of solely \textit{even} modes the leading corrections are quadratic in the magnitude of the force. Indeed, for $m=even$, at second order, using Eq.~\eqref{eq:RG2-0} we get
\begin{align}
\mathcal{R}^2_{E,2}=\frac{8}{N}\sum_{i=odd}\sum_{j=odd}\langle \mathbf{r}_{i}\cdot\mathbf{r}_{j}\rangle^\infty_2
-\frac{\overline{D}_2}{\overline{D}_0}\mathcal{R}^2_{E,0}
 \label{eq:RE2-0}
\end{align}
\begin{figure}
 \includegraphics[width=0.32\textwidth]{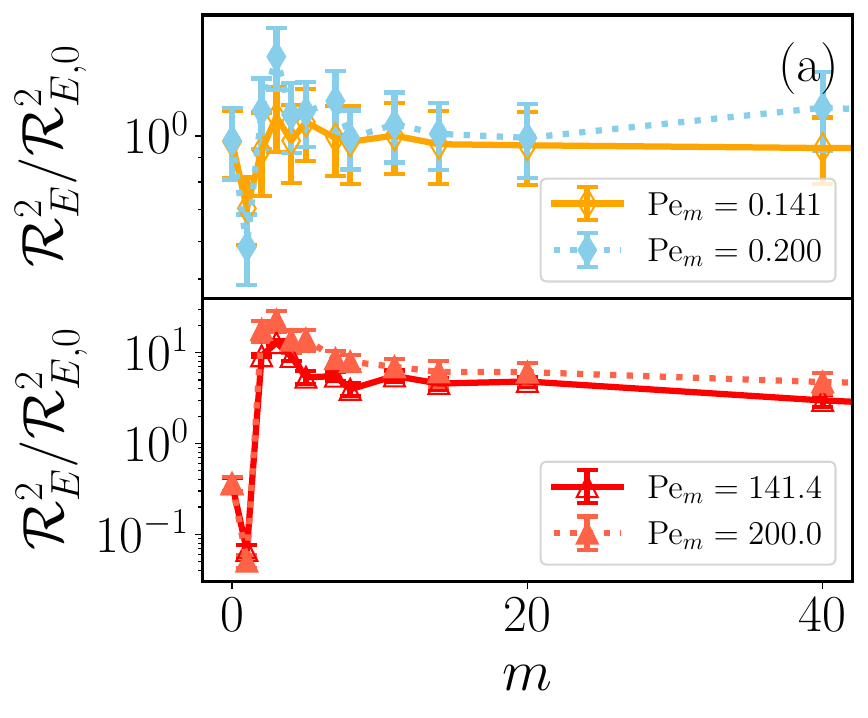}
 \includegraphics[width=0.32\textwidth]{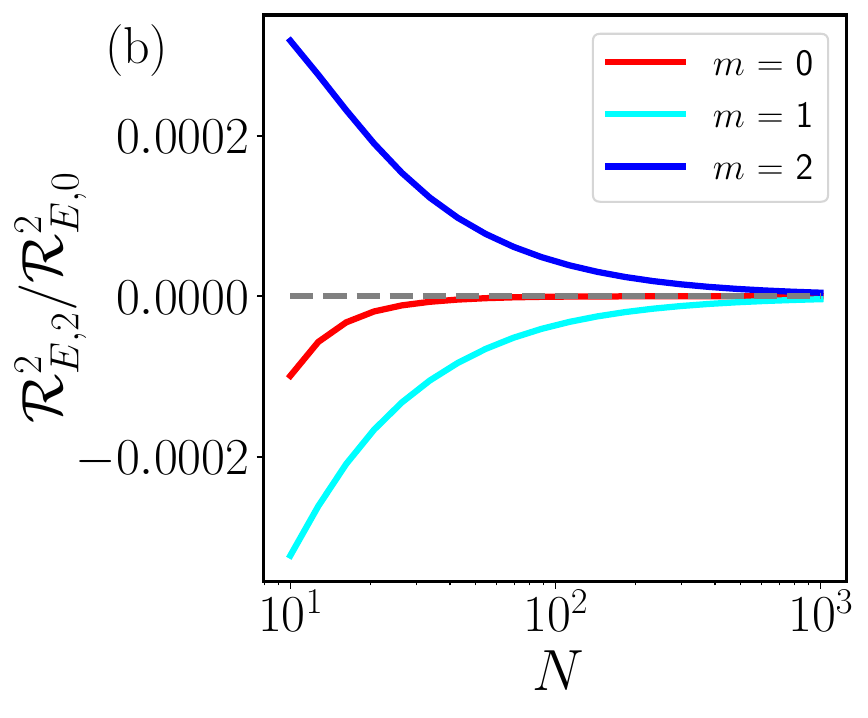}
 \includegraphics[width=0.32\textwidth]{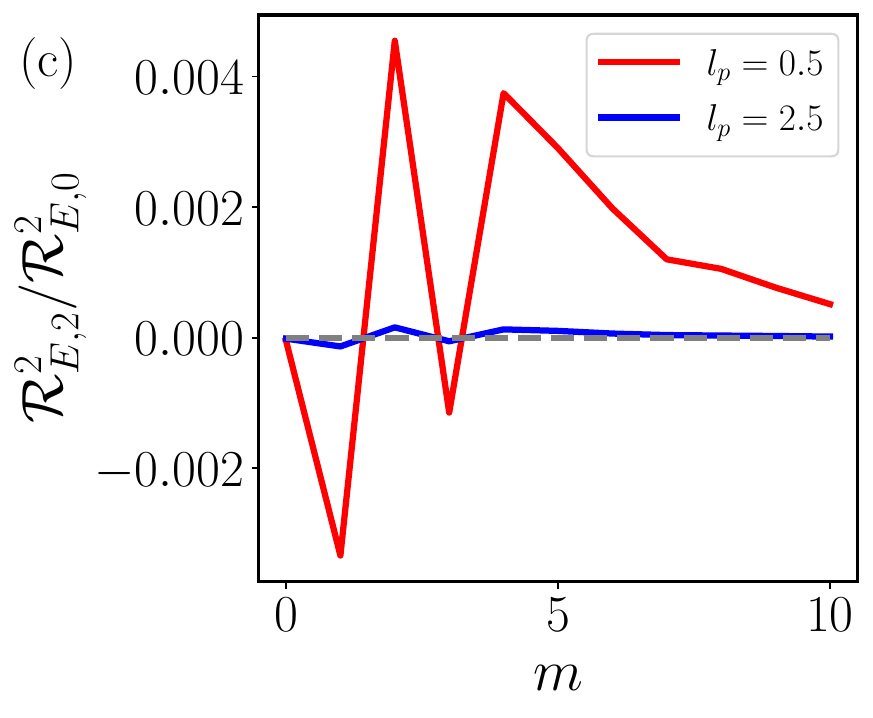}
 \caption{(a) Ratio of the end-to-end square distance, over the \pa{passive} equilibrium value $\mathcal{R}^2_{E}/\mathcal{R}^2_{E,0}$ as a function of the \pa{force} mode $m$ and different values of $\mathrm{Pe}_m$ (with full lines $N=200$ and dotted lines $N=400$) from numerical simulations. (b) \pa{Second-order contribution to the end-to-end square distance} $\mathcal{R}^2_{E,2}$ (Eq.~\eqref{eq:RE2-0}), normalized by $\mathcal{R}^2_{E,0}$, as a function of $N$ for and $l_p=0.5$ and various values of the force mode $m$;  (c) $\mathcal{R}^2_{E,2}/\mathcal{R}^2_{E,0}$ as function of the mode $m$, for $N=100$ and different values of $l_p$.}
 \label{fig:RE2}
\end{figure}
We plot our results in Fig.~\ref{fig:RE2}. Again, we first start from numerical results, plotting the ratio $\mathcal{R}^2_E/\mathcal{R}^2_{E,0}$ as a function of $m$. In both upper and lower sub-panels of Fig.~\ref{fig:RE2}, data is consistent with the previous observations and with the results of Sec.~\ref{sec:rg}. In Fig.~\ref{fig:RE2}b,c, we plot the relative importance of the second-order correction with respect to the zeroth-order one at fixed $\mathrm{Pe}_m = 1$. We remark that these corrections are present for all modes, but they are the {leading order} correction for the \textit{even} modes. We observe the same trends, reported for second order corrections of \pa{the gyration radius,  $\mathcal{R}^2_{G,2}$}: $\mathcal{R}^2_{E,2}$ becomes negligible with increasing $N$ and $m$ and oscillates at small values of $m$, being in particular negative for $m=0$ and $m=1$. However, considering the first order corrections, the theory would predict an overall increase of $\mathcal{R}^2_E$ for $m=1$ and an overall decrease for $m=3$, in contrast to simulation results. This can be possibly due to the fact that we cannot perform simulation at arbitrary small values of $\mathrm{Pe}_m$ and hence higher order corrections can play a role. 
However, more in general, Figs.~\ref{fig:RG} and~\ref{fig:RE2} show that the sign of both $\mathcal{R}^2_{G,2}$ and $\mathcal{R}^2_{E,2}$ changes when crossing a threshold value, $m^*=4$ in the analytical model $m^*=2$ in simulations. For $m\geq m^*$ the chain swells, while the opposite holds for $m<m^*$.

\subsection{Harmonic interactions}

{The shrinking of gyration radius of active polymers upon increasing the active force, in principle, may be regarded as the outcome of an effective attraction among the monomers induced by the activity. Accordingly, in order to check if the behavior of active polymers can be mapped into that of passive polymers in bad solvents (hence experiencing an effective attraction among monomers) we briefly study the case in which there is a net attractive force among the monomers.} 

This amounts to adding an additional contribution
\begin{align}
    F_{i}=-\gamma \sum_{j} (\mathbf{r}_j-\mathbf{r}_i)=-\gamma N \mathbf{r}_{CM}+\gamma N \mathbf{r}_i
\end{align}
where $\mathbf{r}_{CM}=\sum_j \mathbf{r}_j/N$ is the position of the center of mass of the polymer.
Accordingly, in the continuum limit, the equation of motion for the position of each monomer read: 
\begin{equation}
\dot{\mathbf{r}}(s,t)=\mu D\partial_{s}^{2}\mathbf{r}(s,t)+\mu\gamma N \mathbf{r}(s,t)-\mu\gamma N \mathbf{r}_{CM}(t)+\eta(s,t)\label{eq:Rouse2_2}
\end{equation}
Expanding for small values of $\gamma$ we obtain, at $\mathcal{O}(0)$, $\langle \mathbf{r}_i \cdot \mathbf{r}_j \rangle_0=\dfrac{k_BT}{i^2}\delta_{ij}$, while, at $\mathcal{O}(1)$ 
\begin{equation}
       \langle \mathbf{r}_i \cdot \mathbf{r}_j \rangle_1=-\dfrac{1}{D(i^2+j^2)}\left[2\gamma N  \langle \mathbf{r}_i \cdot \mathbf{r}_j \rangle_0 - \gamma N \left(\langle \mathbf{r}_{CM} \cdot \mathbf{r}_j \rangle_0 + \langle \mathbf{r}_i \cdot \mathbf{r}_{CM} \rangle_0\right)\right].
       \label{eq:all-spring}
\end{equation}
For $i=j$ Eq.\eqref{eq:all-spring} reduces to
\begin{align}\label{eq:all-spring-ii}
    \langle \mathbf{r}_i \cdot \mathbf{r}_i \rangle_1=-\dfrac{1}{2D i^2}\left[2\gamma N  \langle \mathbf{r}_i \cdot \mathbf{r}_i \rangle_0-2\gamma N \langle \mathbf{r}_{CM} \cdot \mathbf{r}_i \rangle_0\right]
\end{align}
that for $i\neq 0$ is non vanishing. Moreover, the sign of $\gamma$ is controlling the sign of $\langle \mathbf{r}_i \cdot \mathbf{r}_j \rangle_1$: for $\gamma>0$ (i.e. attractive forces) we have a reduction of the gyration radius (i.e. $\langle \mathbf{r}_i \cdot\mathbf{r}_j \rangle_1<0$) whereas the opposite holds for repulsive forces ($\gamma <0$).
This pinpoints a crucial difference between the collapse due to \textit{non-conservative} or \textit{conservative} forces. Indeed, in a series of experiments where the strength of the force can be tuned, in the first case the gyration radius will have a quadratic dependence on the strength, whereas in the second case a linear dependence. 

\section{Conclusions}

\pa{In this work, we investigated how a nonuniform distribution of tangential activity along the backbone controls the conformations of semiflexible polymers. By combining a perturbative analytical approach with Langevin dynamics simulations, we showed that polymer conformations depend not only on the overall strength of the active forces, but also on the spatial scale over which these forces vary. Patterning the activity along the contour therefore introduces an additional control parameter that has no direct counterpart in passive polymer physics.}

\pa{The main objective of our work is to establish a method to solve the active Rouse model based on a systematic expansion. The result hinges on two fundamental hypotheses: the absence of excluded volume interactions, as usual in Rouse-like models and the presence of a small but finite bending rigidity. This latter ingredient is necessary, in this formulation, to guarantee the convergence of the model, as shown in Ref.~\cite{malgaretti2025coil}. We carry on a faithful comparison between theory and simulations to focus on the insights coming from the analytical solution.}

\pa{Overall, our findings show spatially patterned activity are a mechanism for controlling polymer conformations. By varying the wavelength, orientation, and strength of the activity profile, one can promote compaction, swelling, or conformations that combine global compactness with a large end-to-end extension.}

\pa{In particular, our analysis first identifies the dimensionless parameter $\mathrm{Pe}_m$ (see Eq.~\ref{eq:def-Pe})
as the relevant measure of the departure from equilibrium. This Péclet number compares the work performed by the active force over a distance of the order of the equilibrium polymer size with the thermal energy. Its dependence on $\mathcal{R}_{G,0} = \sqrt{\mathcal{R}^2_{G,0}}$ has an important physical consequence: polymers of different lengths do not respond in the same way to the same local active force. In particular, longer polymers reach the same value of $\mathrm{Pe}_m$ at weaker forcing and are therefore more sensitive to activity. 
Thus, the effect of activity cannot be characterized by an intensive, monomer-level force scale alone, because it also depends on the global size of the object. This size dependence limits the range of the perturbative expansion: increasing the polymer length at fixed force eventually increases $\mathrm{Pe}_m$ beyond the weak-activity regime. Our theory should therefore be understood as an expansion at fixed and sufficiently small $\mathrm{Pe}_m$, rather than as a description of the asymptotic large-$N$ behavior at fixed force.}

\pa{Within this regime, both theory and simulations reveal a crossover between activity-induced shrinking and swelling. Uniform activity and the longest-wavelength modulations, corresponding to the lowest force modes, tend to compact the polymer and reduce its end-to-end distance. For more rapidly varying activity profiles, the chain develops alternating stretched and compressed portions. When sufficiently many such regions are present along the contour, their combined effect produces a globally swollen conformation. The spatial wavelength of the activity thus determines whether active forcing acts predominantly as a source of compaction or extension.}

\pa{The analytical theory predicts a threshold mode \(m^\ast\) separating these two regimes, that is also found in simulations. The precise threshold differs in the two cases, with \(m^\ast=4\) in the perturbative model and \(m^\ast=2\) in the simulations. Quantitative differences are not unexpected, since the theory describes a Gaussian chain in the weak-force limit, whereas the simulations involve finite activity, anharmonic bonds, and microscopic constraints that are absent from the analytical model. Nevertheless, the two approaches agree on the fact that low-mode forcing promotes shrinking, whereas higher-mode forcing produces swelling through the formation of locally stretched segments. }

\pa{A further important result is that activity affects different measures of polymer size in fundamentally different ways. For passive homopolymers, the gyration radius and the end-to-end distance generally provide closely related measures of the spatial extent of the chain. Here, this correspondence breaks down: the gyration radius always responds quadratically to the active force at leading order while the end-to-end distance retains information about the orientation of the force pattern, relative to the two ends of the chain. Indeed, for activity profiles containing odd modes, its leading correction is linear in the force while for even modes is again quadratic. This means that the gyration radius is insensitive to a reversal of the activity profile within the weak-force approximation while the end-to-end distance may reveal it.  }

\pa{This distinction implies that active polymers should not, in general, be described by a single effective size. The same activity profile may reduce the gyration radius while increasing the end-to-end distance. Spatially patterned activity therefore does not merely expand or contract an otherwise self-similar equilibrium conformation; rather, it redistributes extension along the contour and changes the shape of the polymer in a way that different observables probe differently.} 

\pa{The comparison between theory and simulations also indicates where nonlinear effects become relevant. In particular, for the lowest nonuniform mode, the simulations show that reversing the activity profile appears in the gyration radius. This discrepancy suggests that higher-order contributions in $\mathrm{Pe}_m$, which are not included in the present expansion, already become important at the finite activities accessible in the simulations, showing that the low-mode regime is especially sensitive to nonlinearities. This motivates extending the theory beyond the leading perturbative orders.}

\pa{Finally, our results also clarify that activity-induced compaction should not simply be interpreted as an effective attraction between monomers. A genuine conservative attraction changes the gyration radius already linearly in the interaction strength, whereas the leading change produced by tangential activity is quadratic in the force. Although both mechanisms may generate compact conformations, their response to variations of the driving strength is therefore qualitatively different. This distinction could provide a practical way to discriminate between a nonequilibrium collapse driven by propulsion and an equilibrium-like collapse caused by attractive interactions.}

\pa{These results provide a basis for studying more complex patterns of motor activity, stronger nonequilibrium driving, and topologically constrained filaments. Indeed we focused here on a single force mode: however, biological polymers driven by nonuniform distributions of molecular motors and for synthetic active chains whose shape can be controlled through the spatial organization of energy injection. Indeed Eq.~\eqref{eq:f_exp} shows that any force profile that can be expanded on a cosine basis can be included in this formalism. Further, while extremely challenging, the formalism could be extended to semiflexible polymers.}

\pa{Particularly intriguing is the case of topologically constrained filaments. Symmetry and topology place additional constraints on the conformational response: for ring polymers, the absence of ends and the associated mode structure  has well known structural and dynamical consequences in the passive case. 
Activity can still alter ring conformations through higher-order effects, as found in numerical studies~\cite{Locatelli2021, breoni2025effects, miranda2025geometrical}. It will be interesting to better understand which is the leading order of their response.} 

\section{Data availability}
The data are publicly available at \href{https://doi.org/10.5281/zenodo.17953054}{https://doi.org/10.5281/zenodo.17953054}

\bibliography{active_polymer}


\appendix

\section{Details on the analytical expansions}

\subsection{Eigenfunction representation}\label{app:eigen}
In order to get analytical insight into Eq.~\eqref{eq:Rouse}, usually one seeks the eigenfunctions of the operator on the right hand side of Eq.~\eqref{eq:Rouse} that are compatible with the boundary conditions, Eqs.~\eqref{eq:BC-1},\eqref{eq:BC-2}; for the case under study, this is a formidable task. In order to avoid this difficulty, we represent the solution of Eq.\eqref{eq:Rouse} with the \pa{chosen} boundary conditions 
using the eigenfunctions of the passive case, $f(s)=0$. \pa{This} basis is complete and orthonormal~\cite{harnau1995dynamic} \pa{and allows us to write:}
\begin{equation}
\mathbf{r}(s,t) =\sum_{n=0}^{\infty}\mathbf{r}_{n}(t)\psi_n(s) \qquad
f(s) =\sum_{n=0}^{\infty}f_{n}\psi_n(s) \qquad
\boldsymbol{\eta}(s,t) =\sum_{n=0}^{\infty}\boldsymbol{\eta}_{n}(t)\psi_n(s)    
\label{eq:r_fourier}
\end{equation}
where $r_n(t),\eta_n(t)$ are real functions and
\begin{equation}
\left\langle \eta{}_{n}(t)\right\rangle = 0 \qquad 
\left\langle \eta{}_{n,i}(t)\eta{}_{m,j}(t')\right\rangle =  2\mu k_{B}T\delta_{ij}\delta_{n,m}\delta(t-t')
\label{eq:noise_ampl}
\end{equation}
\pa{The representation Eq.\eqref{eq:r_fourier} provides a \textit{normal} modes expansion and the index $n$ refer to such modes.} 
We \pa{also} remark that the inhomogeneous force introduced in Eq.~\eqref{eq:Rouse} leads to a non-trivial representation in the basis of the eigenfunctions of the passive case, as shown in Eq.~\eqref{eq:r_fourier}.  
However, this still represents a convenient basis because, in the expansion that will be performed later, it allows to identify the zeroth order as the passive case, providing a reliable reference.
The eigenfunctions are solution of
\begin{align}
 -\mu G\partial_{s}^{4}\psi_{n}(s)+\mu D\partial_{s}^{2}\psi_n(s)=-\zeta_n \psi_n(s)
\end{align}
and read
\begin{subequations}\label{eq:base}
\begin{align}
 \psi_0&=\sqrt{\frac{1}{N}}\\
 \psi_{n}(s)&=\sqrt{\frac{c_n}{N}}
 \begin{cases}
  k_n\dfrac{\sin(k_n s)}{\cos(k_n N/2)}+\beta_n\dfrac{\sinh(\beta_n s)}{\cosh(\beta_n N/2)}\,, & n \text{ odd}\\
  -k_n\dfrac{\cos(k_n s)}{\sin(k_n N/2)}+\beta_n\dfrac{\cosh(\beta_n s)}{\sinh(\beta_n N/2)}\,, & n \text{ even}\\
 \end{cases}
\end{align}
\end{subequations}
where $c_n$ is the normalization constant and $k_n$ and $\beta_n$ are determined by solving
\begin{align}
 &k_n^3 \sin(k_n N/2)\cosh(\beta_n N/2)-\beta_n^3\cos(k_n N/2)\sinh(\beta_n N/2)-2p(k_n^2+\beta_n^2)\cos(k_n N/2)\cosh(\beta_n N/2)=0\,, n\text{  odd,} \cr
 &k_n^3 \cos(k_n N/2)\sinh(\beta_n N/2)-\beta_n^3\sin(k_n N/2)\cosh(\beta_n N/2)+2p(k_n^2+\beta_n^2)\sin(k_n N/2)\sinh(\beta_n N/2)=0\,, n\text{  even.  } 
\end{align}
Together with the constraint imposed by the boundary conditions $\beta_n^2-k_n^2=\frac{1}{l_p^2}$ the eigenvalues are 
\begin{align}
    {\zeta_n=\mu k_n^2(G k_n^2+D)\,.}
\end{align}
Accordingly, we rewrite Eq.(\ref{eq:Rouse}) in its eigenfunctions representation:
\begin{equation}
\sum_{n=0}^{\infty}\dot{\mathbf{r}}_{n}(t)\psi_n(s)=-\sum_{n=0}^{\infty}\zeta_{n} \mathbf{r}_{n}(t)\psi_n(s)+\dfrac{\sum_{n=0}^{\infty}\sum_{m=0}^{\infty} \mu f_m \mathbf{r}_{n}(t)\psi_m(s)\partial_s\psi_n(s)}{\sqrt{\sum_{m=0}^{\infty} \mathbf{r}_{m}\partial_s \psi_m(s) \cdot\sum_{n=0}^{\infty} \mathbf{r}_{n}\partial_s\psi_n(s)}}+\sum_{n=0}^{\infty}\boldsymbol{\eta}_{n}(t)\psi_n(s)
\end{equation}
We remark that the double sum, over $n$ and $m$ is due to the fact that the force is inhomogeneous and hence it has a non-trivial representation in the basis of the eigenfunctions.
By multiplying both sides by $\psi_i(s)$ and integrating in $ds$ we obtain:
\begin{equation}\label{eq:mode_NL}
\dot{\mathbf{r}}_{i}(t)=-\zeta_i \mathbf{r}_{i}(t)+ \mu \int_{-\frac{N}{2}}^{\frac{N}{2}} \dfrac{\sum_{n=0}^{\infty} \sum_{m=0}^{\infty}  f_m \mathbf{r}_{n}(t)\psi_m(s) \partial_s \psi_n(s)}{\sqrt{\sum_{m=0}^{\infty}  \mathbf{r}_{m}\partial_s \psi_m(s) \cdot\sum_{n=0}^{\infty}  \mathbf{r}_{n}\partial_s\psi_n(s)}} \psi_i(s) ds+\boldsymbol{\eta}_{i}(t)
\end{equation}
In principle, Eqs.~\eqref{eq:mode_NL} can be solved numerically to obtain the amplitudes of the modes from which the conformations of the polymer can be retrieved. However, the magnitude of the modes $\mathbf{r}_i$ are affected by the noise.
Hence typically the focus is on the average values of observables such as the gyration radius or the end-to-end distance which do not depend on the average of the mode amplitudes, rather on their correlations (see Sec.~\ref{sec:rg} and \ref{sec:re}. Accordingly, in the following we derive the steady state values of the correlations among modes which are then required to compute the average values of gyrations radius and end-to-end \pa{distance}.

\subsection{Steady state correlations}\label{sec:ss_corr}

\pa{As mentioned in the introduction, we are interested in deriving closed formulas for the magnitude of the 
the gyration radius, $\mathcal{R}^2_G$ and the end-to-end square distance, $\mathcal{R}^2_E$ 
at steady state (see Sec.\ref{sec:rg} and Sec.\ref{sec:re} for their derivation)
\begin{align}
\mathcal{R}_{G}^2&=\frac{1}{N}\sum_{i\neq 0}{\langle \mathbf{r}_{i}\cdot \mathbf{r}_{i}\rangle^\infty}
\label{app:def-RG}\\
\mathcal{R}^2_E &= \frac{8}{N}\sum_{i=odd}\sum_{j=odd} \langle \mathbf{r}_{i}\cdot\mathbf{r}_{j}\rangle^\infty\label{app:def-REE_0}
\end{align}
which are functions of expectation values of the mode correlations at steady state $\langle \mathbf{r}_i\cdot \mathbf{r}_j\rangle^{\infty}$}
\begin{align}
    \langle \mathbf{r}_i\cdot \mathbf{r}_j \rangle^\infty=\lim\limits_ {t\rightarrow \infty} \langle \mathbf{r}_i(t)\cdot \mathbf{r}_j(t)\rangle\,.
\end{align}
\pa{Note} that we need to compute the cross correlation among modes, since the basis functions we are using, Eqs.~\eqref{eq:base}, are not the eigenfunction of the time evolution operator (i.e., the right hand side of Eq.~\eqref{eq:Rouse}); 
\pa{in practice, we} write first the time evolution of such \pa{correlations} and then we take the limit $t\rightarrow \infty$. Accordingly we \pa{start from}
\begin{equation}
    \partial_t \langle \mathbf{r}_i(t)\cdot \mathbf{r}_j (t)\rangle =  \langle (\partial_t\mathbf{r}_i)\cdot \mathbf{r}_j\rangle+\langle \mathbf{r}_i(t)\cdot (\partial_t\mathbf{r}_j)\rangle
\label{eq:steady_s}
\end{equation}
\pa{The left-hand side,} $\langle \partial_t \mathbf{r}_i(t)\cdot \mathbf{r}_j(t)\rangle$, \pa{reads:}
\begin{align}
\langle \partial_t \mathbf{r}_i(t)\cdot \mathbf{r}_j(t)\rangle&=-\zeta_i \langle \mathbf{r}_{i}(t)\cdot \mathbf{r}_{j}(t)\rangle +\left\langle \mu \int_{-\frac{N}{2}}^{\frac{N}{2}} \dfrac{\sum\limits_{n=0}^{\infty}\sum\limits_{m=0}^{\infty} f_m \psi_m(s) \partial_s\psi_n(s) \, \mathbf{r}_{n}(t)\cdot\mathbf{r}_{j}(t)}{\sqrt{\sum\limits_{m=0}^{\infty} \mathbf{r}_{m} \partial_s\psi_m(s) \cdot\sum\limits_{n=0}^{\infty} \mathbf{r}_{n}\partial_s\psi_n(s)}} \psi_i(s) ds\right\rangle+\langle\boldsymbol{\eta}_{i}(t)\cdot\mathbf{r}_j(t)\rangle.
\label{eq:corr-full}
\end{align}
We approximate the last expression using a mean-field argument, that is, the relevant contributions to the {noise} average of the whole expression come from the {noise} average of the individual correlations; thus Eq.~\eqref{eq:corr-full} can be written as:
\begin{align}
\langle \partial_t \mathbf{r}_i(t)\cdot \mathbf{r}_j(t)\rangle&=-\zeta_i \langle \mathbf{r}_{i}(t) \cdot \mathbf{r}_{j}(t)\rangle +\mu \int_{-\frac{N}{2}}^{\frac{N}{2}} \dfrac{\sum\limits_{n=0}^{\infty}  \sum\limits_{m=0}^{\infty} f_m \psi_m(s) \partial_s\psi_n(s) \, \langle \mathbf{r}_{n}(t)\cdot\mathbf{r}_{j}(t)\rangle}{\sqrt{\sum\limits_{m=0}^{\infty}\sum\limits_{n=0}^{\infty} \partial_s\psi_n(s) \partial_s\psi_m(s) \, \langle \mathbf{r}_{m} \cdot \mathbf{r}_{n}\rangle}} \psi_i(s) ds+\langle\boldsymbol{\eta}_{i}(t)\cdot\mathbf{r}_j(t)\rangle
\label{eq:corr_mf}
\end{align}
One can derive an equivalent expression for  $\langle  \mathbf{r}_i(t)\cdot \partial_t \mathbf{r}_j(t)\rangle$ {by simply interchanging $i$ and $j$ in Eq.~\eqref{eq:corr_mf}}.
{Following Eq.~\eqref{eq:steady_s}, summing $\langle \partial_t \mathbf{r}_i(t)\cdot \mathbf{r}_j(t)\rangle$ (that is, Eq.~\eqref{eq:corr_mf}) and $\langle \mathbf{r}_i(t)\cdot \partial_t \mathbf{r}_j(t)\rangle$ we obtain
\begin{align}
 \partial_t \langle \mathbf{r}_i(t)\cdot \mathbf{r}_j (t)\rangle =&-\zeta_i \langle \mathbf{r}_{i}(t) \cdot \mathbf{r}_{j}(t)\rangle +\mu \int_{-\frac{N}{2}}^{\frac{N}{2}} \dfrac{\sum\limits_{n=0}^{\infty}  \sum\limits_{m=0}^{\infty} f_m \psi_m(s) \partial_s\psi_n(s) \, \langle \mathbf{r}_{n}(t)\cdot\mathbf{r}_{j}(t)\rangle}{\sqrt{\sum\limits_{m=0}^{\infty}\sum\limits_{n=0}^{\infty} \partial_s\psi_n(s) \partial_s\psi_m(s) \, \langle \mathbf{r}_{m} \cdot \mathbf{r}_{n}\rangle}} \psi_i(s) ds+\langle\boldsymbol{\eta}_{i}(t)\cdot\mathbf{r}_j(t)\rangle\nonumber\\
 &-\zeta_j \langle \mathbf{r}_{j}(t) \cdot \mathbf{r}_{i}(t)\rangle +\mu \int_{-\frac{N}{2}}^{\frac{N}{2}} \dfrac{\sum\limits_{n=0}^{\infty}  \sum\limits_{m=0}^{\infty} f_m \psi_m(s) \partial_s\psi_n(s) \, \langle \mathbf{r}_{n}(t)\cdot\mathbf{r}_{i}(t)\rangle}{\sqrt{\sum\limits_{m=0}^{\infty}\sum\limits_{n=0}^{\infty} \partial_s\psi_n(s) \partial_s\psi_m(s) \, \langle \mathbf{r}_{m} \cdot \mathbf{r}_{n}\rangle}} \psi_j(s) ds+\langle\boldsymbol{\eta}_{j}(t)\cdot\mathbf{r}_i(t)\rangle
 \label{eq:ri_rj_time}
\end{align}
By putting to zero the lhs of Eq.~\eqref{eq:ri_rj_time} we can derive closed expressions for $\langle \mathbf{r}_i(t)\cdot \mathbf{r}_j (t)\rangle^\infty$. }
In order to simplify the notation we introduce
\begin{align}\label{eq:def_Theta}
\langle\Theta(s)\rangle=\sum_{m=0}^{\infty}\sum_{n=0}^{\infty} \partial_s\psi_n(s) \partial_s\psi_m(s) \langle\mathbf{r}_{m}\cdot \mathbf{r}_{n}\rangle^\infty
\end{align}
{By substituting Eqs.~\eqref{eq:def_Theta} into Eq.~\eqref{eq:ri_rj_time} and putting the lhs to zero we get} 
\begin{align}
\langle \mathbf{r}_i\cdot \mathbf{r}_j\rangle^\infty=&\frac{\mu}{\Gamma_{i j}}\sum_m f_m \left[
\sum_{n\neq i} \langle \mathbf{r}_{n}\cdot\mathbf{r}_{j}\rangle^\infty
\int_{-\frac{N}{2}}^{\frac{N}{2}} \dfrac{  \psi_m(s) \partial_s\psi_n(s)}{\sqrt{\langle \Theta(s) \rangle}} \psi_i(s) ds+\sum_{n\neq j} \langle \mathbf{r}_{n}\cdot\mathbf{r}_{i}\rangle^\infty \int_{-\frac{N}{2}}^{\frac{N}{2}} \dfrac{   \psi_m(s) \partial_s\psi_n(s)}{\sqrt{\langle \Theta(s) \rangle}} \psi_j(s) ds\right]+\nonumber \\ 
&+\dfrac{ \langle \boldsymbol{\eta}_j\cdot \mathbf{r}_i\rangle^\infty+\langle \boldsymbol{\eta}_i\cdot \mathbf{r}_j\rangle^\infty}{\Gamma_{i j}} \label{eq:rirj}
\end{align}
where
\begin{equation}
\Gamma_{i j} =\zeta_i+\zeta_j-\mu \sum_m f_m\int_{-\frac{N}{2}}^{\frac{N}{2}} \dfrac{\psi_m(s)\psi_i(s)\partial_s\psi_i(s)+\psi_m(s)\psi_j(s)\partial_s\psi_j(s)}{\sqrt{\langle\Theta(s)\rangle}} ds .
\label{eq:def-Gamma}
\end{equation}
Eqs.~\eqref{eq:rirj} are a set of coupled equations for the correlations among the different modes. 
\subsection{Analytical Expansions I: small active forces}\label{app:1}

Our strategy for tackling Eq.~\eqref{eq:rirj} analytically consists of a systematic expansion of all modes at small values of $f_m$. 
In order to calculate $\langle \mathbf{r}_i(t)\cdot \mathbf{r}_j\rangle$ at the different orders in $f_m$ we need to calculate $\langle \mathbf{r}_i(t)\cdot \boldsymbol{\eta}_j\rangle$. We will employ the formal solution of $\mathbf{r}_i(t)$, obtained  integrating Eq.~\eqref{eq:Rouse} in its eigenfunction representation: it reads:
\begin{align}
\mathbf{r}_i(t)=\int_0^te^{-\frac{\Gamma_{ii}}{2} (t-\tau)}\left[\mu \sum_m f_m\int_{-\frac{N}{2}}^{\frac{N}{2}} \dfrac{\sum\limits_{n\neq i}  \psi_m(s) \partial_s\psi_n(s) \mathbf{r}_{n}(\tau)}{\sqrt{\langle\Theta(s)\rangle}} \psi_i(s) ds+\boldsymbol{\eta}_{i}(\tau)\right]d\tau.
\label{eq:form-sol}
\end{align}
Notice that, at this stage, we consider a generic active force, represented by its expansion Eq.~\eqref{eq:f_exp}. As such, these results are valid for any inhomogeneous force, whose functional form is sufficiently regular along the polymer backbone; at a certain stage, we will fix a value of the ``force mode'' $m$ to consider single-mode active forces. 
In order to simplify the notation we introduce the stead state expectation value expansion of $\langle \Theta(s)\rangle$ up to second order in $f_m$
\begin{align}
 \langle \Theta(s)\rangle=\langle \Theta(s)\rangle_0+\langle \Theta(s)\rangle_1+\langle \Theta(s) \rangle_2.
 \label{eq:theta-exp}
\end{align}

\paragraph{Zeroth order} 

At zeroth order in $f_m$, using Eqs.~\eqref{eq:noise_ampl},~\eqref{eq:form-sol}, we obtain:
\begin{align}
   \langle \boldsymbol{\eta}_j\cdot \mathbf{r}_i\rangle^\infty_0=\langle \boldsymbol{\eta}_j\cdot \mathbf{r}^{(0)}_i\rangle^\infty=\lim_{t\rightarrow \infty} \int_0^t e^{-\frac{\Gamma_{ii}^{(0)}}{2} (t-\tau)}\langle \boldsymbol{\eta}_i(\tau)\cdot \boldsymbol{\eta}_j(t)\rangle d\tau=\delta_{ij}2d\mu k_BT
   \label{eq:app_r-eta-0}
\end{align}
where $\mathbf{r}^{(0)}_i(t)$ is the zeroth-order solution obtained from Eq.~\eqref{eq:form-sol}  and $d$ is the dimension of the space in which the polymer is embedded in. Hence we have
\begin{align}
    \langle \mathbf{r}_i\cdot \mathbf{r}_j\rangle^\infty_0=\lim_{t\rightarrow \infty} \int_0^t\int_0^te^{-\frac{\Gamma^{(0)}_{ii}}{2} (t-\tau_1)}e^{-\frac{\Gamma_{jj}^{(0)}}{2} (t-\tau)}\delta_{ij}\langle \boldsymbol{\eta}_i(\tau_1)\cdot \boldsymbol{\eta}_j(\tau)\rangle d\tau d\tau_1=\delta_{ij}d\dfrac{\mu k_BT}{\zeta_i}
    \label{eq:app_corr_ii}
\end{align}
where $\Gamma^{(0)}_{ii}=2\zeta_i = 2\mu k_i^2(G k_i^2+D)$ is the zeroth order contribution in $f_m$ to $\Gamma_{ii}$. For later use we note that
\begin{align}
 \langle\Theta(s)\rangle_0=\sum_n \sum_m \partial_s \psi_n(s) \partial_s \psi_m(s) \langle \mathbf{r}_n\cdot \mathbf{r}_{m}\rangle^\infty_0=\sum_n (\partial_s \psi_n(s))^2 \langle \mathbf{r}_n(t)\cdot \mathbf{r}_n(t)\rangle_0^{\infty}
\end{align}
as, at this order, all non-diagonal correlators vanish.
Hence, since all $\psi(s)$ have a definite parity then $\langle\Theta(s,t)\rangle_0$ is an even function in the interval $-N/2<s<N/2$.\\

\paragraph{First order} 
At first order, 
using the same mean-field approximation introduced in Sec.~\ref{sec:ss_corr} we have
\begin{align}
\left\langle \boldsymbol{\eta}_j(t)\cdot \mathbf{r}_i(t)\right\rangle_1=&\mu \sum_m f_m \int_{-\frac{N}{2}}^{\frac{N}{2}}\int_0^t \sum_{n\neq i} e^{-\frac{\Gamma_{nn}^{(0)}}{2} (t-\tau)}
 \dfrac{ \psi_m(s)\psi_i(s)\partial_s\psi_n(s)}{\sqrt{\langle\Theta(s)\rangle_0}}\langle\boldsymbol{\eta}_j(t)\cdot \mathbf{r}_{n}(\tau) \rangle_0 d\tau ds+\nonumber\\
 &-\int_0^te^{-\frac{\Gamma_{ii}^{(0)}}{2} (t-\tau)} \dfrac{\Gamma_{i i}^{(1)}(t-\tau)}{\Gamma_{i i}^{(0)}} \langle\boldsymbol{\eta}_j(t)\cdot \boldsymbol{\eta}_i(\tau) \rangle_0 d\tau
 \label{eq:corr_ii-1}
\end{align}
where
\begin{align}
  \Gamma_{ij}^{(1)}&=-\mu \sum_m f_m\int_{-\frac{N}{2}}^{\frac{N}{2}} \frac{\psi_m(s)\psi_i(s)\partial_s\psi_i(s)+\psi_m(s)\psi_j(s)\partial_s\psi_j(s)}{\sqrt{\langle \Theta(s,t)\rangle_0}}ds\nonumber\\
\end{align}
is the first order contribution of $\Gamma_{ij}$ (see Eq.~\eqref{eq:def-Gamma}).
We highlight that the following equalities hold at $t \to \infty$:
\begin{equation}
 \int_0^t e^{-\frac{\Gamma_{nn}^{(0)}}{2} (t-\tau)} \langle\boldsymbol{\eta}_j(t)\cdot \mathbf{r}_{n}(\tau) \rangle_0 d\tau =\int_0^t e^{-\frac{\Gamma_{nn}^{(0)}}{2} (t-\tau)}\int_0^\tau e^{-\frac{\Gamma_{nn}^{(0)}}{2} (\tau-\tau_1)} \langle\boldsymbol{\eta}_j(t)\cdot \boldsymbol{\eta}_n(\tau_1) \rangle_0 \, d\tau_1 d\tau=0 
 \label{eq:double-delta}
\end{equation}
where we used the result Eq.~\eqref{eq:app_r-eta-0}, and
\begin{equation}
 \int_0^t e^{-\frac{\Gamma_{nn}^{(0)}}{2} (t-\tau)} \dfrac{\Gamma_{i i}^{(1)}(t-\tau)}{\Gamma_{i i}^{(0)}} \langle\boldsymbol{\eta}_j(t)\cdot \boldsymbol{\eta}_i(\tau) \rangle_0 d \tau =0;
 \label{eq:eta-eta_1}
\end{equation}
{both equalities are using the fact that $\boldsymbol{\eta}$ is a white noise.} Using these results Eq.~\eqref{eq:corr_ii-1} reads
 \begin{align}
  \langle \boldsymbol{\eta}_j(t)\cdot \mathbf{r}_i(t)\rangle_1=\langle \boldsymbol{\eta}_i(t)\cdot \mathbf{r}_j(t)\rangle_1=0
  \label{eq:eta-eta-1}
 \end{align}
%
from which we obtain
\begin{align}
\langle \mathbf{r}_i\cdot \mathbf{r}_j\rangle^\infty_1=&\mu \sum_m f_m \int_{-\frac{N}{2}}^{\frac{N}{2}}\psi_m(s)\dfrac{\sum_{n\neq i}\psi_i(k_i s)\partial_s\psi_n(s)\langle \mathbf{r}_n\cdot \mathbf{r}_j\rangle^\infty_0+\sum_{n\neq j}\psi_j(s)\partial_s\psi_n(s) \langle \mathbf{r}_i\cdot \mathbf{r}_n\rangle^\infty_0}{\Gamma_{i j}^{(0)}\sqrt{\langle\Theta(s)\rangle_0}}ds \nonumber\\
=&\frac{\mu}{\Gamma_{ij}^{(0)}}(1-\delta_{ij})d\sum_m f_m \left[\frac{\mu k_BT}{\zeta_j}\xi_{ijm}^{(0)}+\frac{\mu k_BT}{\zeta_i}\xi_{jim}^{(0)}\right]
\label{eq:o_1-1}
\end{align}
where the term $1-\delta_{ij}$ stems from the constraints on the sums. The last expression shows that under the assumption of small active forces we have that 
\begin{align}
    \langle \mathbf{r}_i\cdot \mathbf{r}_i\rangle^\infty_1= 0
\end{align}
i.e., the diagonal elements are identically zero.
We used Eq.~\eqref{eq:app_corr_ii} to obtain the second line; we introduce  
\begin{equation}
 \xi_{ijm}^{(0)}= \int_{-\frac{N}{2}}^{\frac{N}{2}}\dfrac{\psi_m(s)\psi_i(s)\partial_s\psi_j(s)}{\sqrt{\langle\Theta(s)\rangle_0}}ds
 \label{eq:xi-0}
\end{equation}

\paragraph{Second order}
At second order,  
using again the same mean-field approximation we obtain
\begin{align}\label{eq:o_2-1}
 \left\langle \boldsymbol{\eta}_j(t)\cdot \mathbf{r}^{(2)}_i(t)\right\rangle=& \int_0^t \mu \sum_m f_m\int_{-\frac{N}{2}}^{\frac{N}{2}}\!\!e^{-\frac{\Gamma_{ii}^{(0)}}{2} (t-\tau)}
\left[\dfrac{ \sum_{n\neq i} \left\langle \boldsymbol{\eta}_j(t) \cdot \mathbf{r}^{(1)}_{n}(\tau)\right\rangle\psi_m(s)\psi_i(s)\partial_s\psi_n(s)}{\sqrt{\langle\Theta(s)\rangle_0}}\right]ds d\tau  +\nonumber\\
&- \int_0^t \frac{\mu}{2} \sum_{m} f_m \int_{-\frac{N}{2}}^{\frac{N}{2}}\!\!e^{-\frac{\Gamma_{ii}^{(0)}}{2} (t-\tau)}
\left[\dfrac{ \sum_{n\neq i} \left\langle \boldsymbol{\eta}_j(t) \cdot \mathbf{r}_{n}(\tau)\right\rangle_0\psi_m(s)\psi_i(s)\partial_s\psi_n(s)}{\sqrt{\langle\Theta(s)\rangle_0}}\dfrac{\langle\Theta(s)\rangle_1}{\langle\Theta(s)\rangle_0}\right]ds d\tau+\nonumber\\
&+\int_0^t \frac{\mu}{2} \sum_{m} f_m \int_{-\frac{N}{2}}^{\frac{N}{2}}\!\!e^{-\frac{\Gamma_{ii}^{(0)}}{2} (t-\tau)}\dfrac{\Gamma_{ii}^{(1)}}{\Gamma_{ii}^{(0)}}(t-\tau)
\left[\dfrac{ \sum_{n\neq i} \left\langle \boldsymbol{\eta}_j(t) \cdot \mathbf{r}^{(0)}_{n}(\tau)\right\rangle\psi_m(s)\psi_i(s)\partial_s\psi_n(s)}{\sqrt{\langle\Theta(s)\rangle_0}}\right]ds d\tau+\nonumber\\
&+\int_0^t e^{-\frac{\Gamma_{ii}^{(0)}}{2} (t-\tau)}\dfrac{\Gamma_{ii}^{(2)}}{\Gamma_{ii}^{(0)}}(t-\tau)\left\langle \boldsymbol{\eta}_j(t) \cdot \boldsymbol{\eta}_i(\tau)\right\rangle_0 d\tau\cr
=&0 
\end{align}
We note that the first and second terms of Eq.~\eqref{eq:o_2-1} vanish due to Eq.~\eqref{eq:double-delta}, the third and fourth terms of  Eq.~\eqref{eq:o_2-1} vanish due to Eq.~\eqref{eq:eta-eta_1}. 
Finally, using Eq.~\eqref{eq:eta-eta-1} we have:
\begin{align}\label{eq:rr2}
\left\langle \mathbf{r}_i\cdot \mathbf{r}_j\right\rangle^\infty_2=&\frac{\mu }{\Gamma_{i j}^{(0)}}\sum_m f_m\left[\sum_{n\neq i} \left\langle \mathbf{r}_j\cdot\mathbf{r}_{n}\right\rangle^\infty_1 \xi^{(0)}_{inm}+\sum_{n\neq j} \left\langle \mathbf{r}_i\cdot\mathbf{r}_{n}\right\rangle^\infty_1 \xi^{(0)}_{jnm}\right] + \nonumber\\
&-\frac{\mu }{\Gamma_{i j}^{(0)}}\sum_m f_m\left[\sum_{n\neq i} \left\langle \mathbf{r}_j\cdot\mathbf{r}_{n}\right\rangle^\infty_0 \xi^{(0)}_{inm}+\sum_{n\neq j} \left\langle \mathbf{r}_i\cdot\mathbf{r}_{n}\right\rangle^\infty_0 \xi^{(0)}_{jnm}\right]\frac{\Gamma^{(1)}_{ij}}{\Gamma^{(0)}_{ij}}+ \nonumber\\
&+\frac{\mu }{2\Gamma_{i j}^{(0)}}\sum_m f_m\left[\sum_{n\neq i} \left\langle \mathbf{r}_j\cdot\mathbf{r}_{n}\right\rangle^\infty_0 \xi^{(1)}_{inm}+\sum_{n\neq j} \left\langle \mathbf{r}_i\cdot\mathbf{r}_{n}\right\rangle^\infty_0 \xi^{(1)}_{jnm}\right] +\nonumber\\
&-\frac{1}{2\Gamma_{i j}^{(0)}}\left[
\dfrac{\Gamma_{i j}^{(2)}}{\Gamma_{i j}^{(0)}}\left(\left\langle \boldsymbol{\eta}_j\cdot \boldsymbol{\eta}_i\right\rangle^\infty_0+\left\langle \boldsymbol{\eta}_i\cdot \boldsymbol{\eta}_j\right\rangle^\infty_0\right)\right]
\end{align}
where
\begin{align}
 \Gamma_{i j}^{(2)}=&-\mu \sum_m f_m \left(\xi^{(1)}_{iim}+\xi^{(1)}_{jjm}\right)\label{eq:Gamma-2}\\
 \xi^{(1)}_{ijm}=&\int_{-\frac{N}{2}}^{\frac{N}{2}}\dfrac{\psi_m(s)\psi_i(s)\partial_s\psi_j(s)}{\sqrt{\langle\Theta(s)\rangle_0}}\dfrac{\langle\Theta(s)\rangle_1}{\langle\Theta(s)\rangle_0}  ds\label{eq:xi-1}
\end{align}
Using Eqs.~\eqref{eq:noise_ampl},~\eqref{eq:app_corr_ii} we obtain:
\begin{align}
\left\langle \mathbf{r}_i\cdot \mathbf{r}_j\right\rangle^\infty_2=&\frac{\mu }{\Gamma_{i j}^{(0)}}\sum_m f_m\left[\sum_{n\neq i} \left\langle \mathbf{r}_j\cdot\mathbf{r}_{n}\right\rangle^\infty_1 \xi^{(0)}_{inm}+\sum_{n\neq j} \left\langle \mathbf{r}_i\cdot\mathbf{r}_{n}\right\rangle^\infty_1 \xi^{(0)}_{jnm}\right]+\nonumber\\
&+(1-\delta_{ij})\frac{d\mu^2 k_BT }{2\Gamma_{i j}^{(0)}}\sum_m f_m\left[\frac{\xi^{(1)}_{ijm}}{\zeta_j}+\frac{\xi^{(1)}_{jim}}{\zeta_i}\right]-d\mu k_BT \delta_{ij}\dfrac{ \Gamma_{i j}^{(2)}}{\left(\Gamma_{i j}^{(0)}\right)^2}
\label{eq:rr-2}
\end{align}
where the $1-\delta_{ij}$ stems from the restriction on the sum on the second line of Eq.\eqref{eq:rr2}.

\subsection{Analytical Expansion II: Short persistence length}
\label{sec:expansion_lp}

\pa{The calculations and the expressions can be further simplified if the persistence length is much shorter then the polymer length, that is, $l_p\ll N\sigma$. This case is not only convenient from the technical perspective, but it is also most relevant from the physical one since flexible polymers are, naturally, more affected by activity with respect to their rigid counterparts, especially from the perspective of the polymer conformations.}
In this regime, the relation $\beta^2_n=k^2_n+1/l^2_p \gg N$ holds and, accordingly, the hyperbolic functions are of order unity only in the vicinity of the edges of the polymer. In such a regime, we can approximate $k_n=\frac{\pi n}{N}$ and take $\beta_n\rightarrow \infty$ and hence the eigenfunction can be approximated by their trigonometric contribution.
In order to simplify the notation, it is more convenient to map the position along the backbone, $s$, on $s\in[0:N]$ rather than $s\in[-N/2:N/2]$; the boundary conditions (Eqs.\eqref{eq:BC-1},~\eqref{eq:BC-2}) are then evaluated at $s=0,N$ and the eigenfunctions read:
\begin{subequations}\label{eq:simple_base}
\begin{align}
 \psi_0&=\sqrt{\frac{1}{N}}\\
 \psi_n(s)&=\sqrt{\frac{2}{N}}\cos(k_n s)
\end{align}
\end{subequations}
Further, we rewrite the eigenvalues as
\begin{equation}
 \zeta_i =\mu k_i^2(G k_i^2 +D)=\mu i^2 Q (\Lambda_p i^2 +1), \qquad \text{where} \qquad Q =\frac{D\pi^2}{N^2}, \qquad \Lambda_p =\frac{l^2_p \pi^2}{N^2}
 \label{eq:def-Lambda}
\end{equation}
and we represent the external force as 
\begin{equation}
    f(s)=\sum_{m} f_{m} \cos(k_{m} s).
    \label{eq:f_exp}
\end{equation}
\pa{The representation of the force profile in Eq.~\eqref{eq:f_exp} is valid for any function that can be expanded on a cosine basis, that matches the expansion basis Eq.~\eqref{eq:simple_base}.}

Shifting the domain of definition of the variable $s$, we obtain the eigenfunctions Eqs.~\ref{eq:simple_base}. 
Moreover we rewrite the eigenvalues as
\begin{equation}
 \zeta_i =\mu k_i^2(G k_i^2 +D)=\mu i^2 \frac{D\pi^2}{N^2}(\Lambda_p i^2 +1), \qquad 
 \Lambda_p =\frac{l^2_p \pi^2}{N^2}\label{eq:app_def-Lambda}
\end{equation}

\paragraph{Zeroth order}
Defining $\bar{z}_n=n^2(1+\Lambda_p n^2)$ and using the definitions of $Q$ and $\Lambda_p$, Eqs.~\eqref{eq:app_def-Lambda} at zeroth order we have:
\begin{align}
\Gamma_{ij}^{(0)} &= \frac{\pi^2 \mu D}{N^2}(\bar{z}_i+\bar{z}_j)\label{eq:G0}\\
\langle \mathbf{r}_i\cdot \mathbf{r}_j\rangle^\infty_0 &= \delta_{ij}d\frac{k_BT N^2}{\pi^2 D}\frac{1}{\bar{z}_i}\label{eq:a_rirj0}\\
\langle \Theta(s)\rangle_0 &= 2d\frac{k_BT}{N D}\sum_{n} n^2\dfrac{\sin^2(k_n s)}{\bar{z}_n}\label{eq:Theta_0-approx}
\end{align}
Eq.~\eqref{eq:a_rirj0} corresponds to Eq.~\eqref{eq:rr0-final} of the main text. The expression in Eq.~\eqref{eq:Theta_0-approx} can be rewritten in terms of hypergeometric functions. Numerical inspection of Eq.~\eqref{eq:Theta_0-approx} in the regime $\Lambda_p \ll N$ shows that
\begin{align}
 \sum_{n=0}^{\infty}n^2\frac{\sin^2(k_n s)}{\bar{z}_n}\simeq \frac{\pi}{4}\frac{1}{\sqrt{\Lambda_p}} \dfrac{\cosh[\kappa N/2]-\cosh[\kappa (s-N/2)]}{\cosh[\kappa N/2]-1}
\end{align}
with $\kappa\simeq \frac{1}{N}{2\pi }/{\sqrt{\Lambda_p}}\equiv {2}/{l_p}$; when $\kappa N\gg 1$ we can approximate
\begin{align}
 \sum_{n=0}^{\infty}n^2\frac{\sin^2(k_n s)}{\bar{z}_n}\simeq 
 \frac{\pi}{4}\frac{1}{\sqrt{\Lambda_p}}
 \label{eq:chi-approx}
\end{align}
i.e., with the value it attains at its maximum. Accordingly, using Eq.~\eqref{eq:def-D} we have
\begin{align}
 \langle\Theta(s)\rangle_0&\simeq \frac{d}{2}\frac{k_BT}{D l_p}=b^2\label{eq:Theta0_app-1}\\
\xi_{i{j}m}^{(0)}&\simeq -\left(\frac{2}{N}\right)^\frac{3}{2}\frac{1}{b}\int_{-N/2}^{N/2} \frac{\pi j}{N} \sin(k_{j} s)\cos(k_i s)\cos(k_m s)ds \simeq -\left(\frac{2}{N}\right)^\frac{3}{2} \frac{1}{b}\bar\xi_{ijm}^{(0)}\label{eq:xi_0_approx}
\end{align}
where we defined
\begin{align}
 \bar{\xi}_{ijm}^{(0)}&=\frac{\pi j}{N}\int_{-N/2}^{N/2}  \sin( k_{j} s)\cos(k_i s)\cos(k_m s)ds\nonumber\\
 &=\frac{j^2  \left((-1)^{i+m+j}-1\right) \left(i^2+m^2-j^2\right)}{  (i-m-j) (i+m-j) (i-m+j) (i+m+j)}
 \label{eq:xi_0_approx-1}
\end{align}

\paragraph{First order}
Using the definitions of $Q$ and $\Lambda_p$, Eqs.~\eqref{eq:app_def-Lambda} together with ~\eqref{eq:G0},~\eqref{eq:xi_0_approx} Eq.~\eqref{eq:o_1-1},~\eqref{eq:def-Theta1} and Eq.~\eqref{eq:xi-1} respectively read
\begin{align}
 \langle \mathbf{r}_{i}\cdot \mathbf{r}_{j}\rangle^\infty_1&=-8\sqrt{2}\frac{b^2 N^2}{\pi^4 d^2}\frac{1}{l_p^2} \frac{(1-\delta_{ij})}{\bar{z}_{i}+\bar{z}_{j}}\sum_{m} \frac{f_m b \sqrt{N}}{k_BT}\left(\frac{\bar{\xi}_{ijm}^{(0)}}{\bar{z}_{j}}+\frac{\bar{\xi}_{jim}^{(0)}}{\bar{z}_{i}}\right)\label{eq:rr1}\\
 \langle\Theta(s)\rangle_1 = &\frac{2}{N}\sum_{i}\sum_{j} k_{i} k_{j} \sin(k_{i} s) \sin(k_{j} s)\langle \mathbf{r}_{i}\cdot \mathbf{r}_{j}\rangle^\infty_1\label{eq:def-Theta1}
\end{align}
and hence 
\begin{align}
    \xi_{ijm}^{(1)}&\simeq \left(\frac{2}{N}\right)^\frac{5}{2}\frac{1}{b^3}\sum_{l}\sum_{q} \langle \mathbf{r}_{l}\cdot \mathbf{r}_{q}\rangle^\infty_1 \int_{-N/2}^{N/2} \frac{\pi^3 j\cdot l\cdot q}{N^3} \cos(k_r s)\cos(k_i s)\sin(k_j s) \sin(k_{l} s)  \sin(k_{q} s)  ds \\
    &\simeq -\frac{4}{\pi^2 l_p^2}\frac{1}{N^\frac{5}{2}}\frac{1}{b}\sum_{l}\sum_{q} (1-\delta_{lq}) \frac{j\cdot l\cdot q}{\bar{z}_{l}+\bar{z}_{q}}\sum_{r} \frac{f_{r} b \sqrt{N}}{k_BT}\left(\frac{\bar{\xi}_{lqr}^{(0)}}{\bar{z}_{q}}+\frac{\bar{\xi}_{qlr}^{(0)}}{\bar{z}_{l}}\right)A_{ijqlm}\\
    &\simeq -\frac{4}{\pi^2 l_p^2}\frac{1}{N^\frac{5}{2}}\frac{1}{b} \bar\xi_{ijm}^{(1)}
\end{align}
with
\begin{align}
 \bar{\xi}_{ijm}^{(1)}&= \sum_{r} \frac{f_{r} b \sqrt{N}}{k_BT} \sum_{l}\sum_{q} (1-\delta_{lq}) \frac{j\cdot l\cdot q}{\bar{z}_{l}+\bar{z}_{q}}\left(\frac{\bar{\xi}_{lqr}^{(0)}}{\bar{z}_{q}}+\frac{\bar{\xi}_{qlr}^{(0)}}{\bar{z}_{l}}\right)A_{ijqlm}\\
 A_{ijqlm}&=\frac{16\pi}{N}\int_{-N/2}^{N/2} \cos(k_{m} s)\cos(k_i s)\sin(k_j s) \sin(k_{l} s)  \sin(k_{q} s)  ds
 \end{align}

\paragraph{Second order}
At second order, Eq.~\eqref{eq:rr-2} reads:
\begin{align}
\left\langle \mathbf{r}_i\cdot \mathbf{r}_j\right\rangle^\infty_2=&\frac{64}{\pi^6}\frac{b^2 N^2}{l_p d^3}\frac{1}{\bar{z}_i+\bar{z}_j}\sum_m \frac{f_m b \sqrt{N}}{k_BT}\sum_l \frac{f_l b \sqrt{N}}{k_BT}\left[\sum_{n\neq i} \frac{1-\delta_{jn}}{\bar{z}_n+\bar{z}_j}\left(\frac{\bar{\xi}^{(0)}_{jnl}}{\bar{z}_n}+\frac{\bar{\xi}^{(0)}_{njl}}{\bar{z}_j}\right) \bar{\xi}^{(0)}_{inm}+\sum_{n\neq j} \frac{1-\delta_{in}}{\bar{z}_n+\bar{z}_i}\left(\frac{\bar{\xi}^{(0)}_{inl}}{\bar{z}_n}+\frac{\bar{\xi}^{(0)}_{nil}}{\bar{z}_i}\right) \bar{\xi}^{(0)}_{jnm}\right]+\nonumber\\
&-\frac{8 b^2 N^2}{d \pi^6}\frac{1-\delta_{ij}}{\bar{z}_i+\bar{z}_j}\sum_{m} \frac{f_{m} b \sqrt{N}}{k_BT}\left[\frac{\bar{\xi}^{(1)}_{ijm}}{\bar{z}_j}+\frac{\bar{\xi}^{(1)}_{jim}}{\bar{z}_i}\right]-\delta_{ij}\frac{8 b^2 N^2}{\pi^6 d}\frac{1}{\bar{z}_i^2}\sum_{m} \frac{f_{m} b \sqrt{N}}{k_BT} \bar{\xi}^{(1)}_{iim}
\label{eq:rr-2-app}
\end{align}
That for diagonal terms becomes
\begin{align}
\left\langle \mathbf{r}_i\cdot \mathbf{r}_{i}\right\rangle^\infty_2=&\frac{128}{\pi^6}\frac{b^2 N^2}{l_p d^3}\frac{1}{\bar{z}_i}\sum_m \frac{f_m b \sqrt{N}}{k_BT}\sum_l \frac{f_l b \sqrt{N}}{k_BT}\left[\sum_{n\neq i} \frac{1-\delta_{in}}{\bar{z}_n+\bar{z}_i}\left(\frac{\bar{\xi}^{(0)}_{inl}}{\bar{z}_n}+\frac{\bar{\xi}^{(0)}_{nil}}{\bar{z}_i}\right) \bar{\xi}^{(0)}_{inm}\right]-\frac{8 b^2 N^2}{\pi^6 d}\frac{1}{\bar{z}_i^2}\sum_{m} \frac{f_{m} b \sqrt{N}}{k_BT} \bar{\xi}^{(1)}_{iim}
\label{eq:app-r2}
\end{align}

\subsection{Single force mode and relevant P\'eclet number}\label{app:single-mode}

The inhomogeneity of the force introduced in Eq.~\eqref{eq:Rouse} leads to the sums in Eqs.~\eqref{eq:rr1},\eqref{eq:app-r2}. In order to get analytical insight and to reduce the number of parameters in the following we focus on the case in which the amplitudes of the modes of the force, introduced in Eq.~\eqref{eq:r_fourier}, are all equal to zero but one; accordingly, 
\begin{equation}
    f(s)= f_{m} \cos(k_{m} s).
    \label{eq:f_single}
\end{equation}
where $m$ is the force mode. In this case, by comparing the values of the correlation at zeroth order to those at first order we obtain that the latter are much smaller then the former when
\begin{align}
    \mathrm{Pe}_m = \frac{f_m b \sqrt{N}}{k_BT} \ll 1\, ,
    \label{eq:app_def-Pe}
\end{align}
which indeed has the form of a P\'eclet number. Is interesting to note that the P\'eclet number in Eq.~\eqref{eq:app_def-Pe} scales as $\sqrt{N}$. 
For such a single-mode force, we can simplify the expression at first order 
\begin{align}
\langle \mathbf{r}_{i}\cdot \mathbf{r}_{j}\rangle^\infty_1&=-8\sqrt{2}\frac{b^2 N^2}{\pi^4 d^2}\frac{1}{l_p^2} \frac{(1-\delta_{{ij}})}{\bar{z}_{i}+\bar{z}_{j}} \sum_{m} \mathrm{Pe}_m\left(\frac{\bar{\xi}_{{ijm}}^{(0)}}{\bar{z}_{j}}+\frac{\bar{\xi}_{{jim}}^{(0)}}{\bar{z}_{i}}\right),
 \label{eq:rr1-Pe}
 \end{align} 
as well as at second order 
\begin{align}
\left\langle \mathbf{r}_i\cdot \mathbf{r}_j\right\rangle^\infty_2=&\frac{64}{\pi^6}\frac{b^2 N^2}{l_p d^3}\frac{1}{\bar{z}_i+\bar{z}_j} \mathrm{Pe}_m^2\left[\sum_{n\neq i} \frac{1-\delta_{jn}}{\bar{z}_n+\bar{z}_j}\left(\frac{\bar{\xi}^{(0)}_{jnm}}{\bar{z}_n}+\frac{\bar{\xi}^{(0)}_{njm}}{\bar{z}_j}\right) \bar{\xi}^{(0)}_{inm}+\sum_{n\neq j} \frac{1-\delta_{in}}{\bar{z}_n+\bar{z}_i}\left(\frac{\bar{\xi}^{(0)}_{inm}}{\bar{z}_n}+\frac{\bar{\xi}^{(0)}_{nim}}{\bar{z}_i}\right) \bar{\xi}^{(0)}_{jnm}\right]+\nonumber\\
&-\frac{8 b^2 N^2}{d \pi^6}\frac{1-\delta_{ij}}{\bar{z}_i+\bar{z}_j} \mathrm{Pe}_m\left[\frac{\bar{\xi}^{(1)}_{ijm}}{\bar{z}_j}+\frac{\bar{\xi}^{(1)}_{jim}}{\bar{z}_i}\right]-\delta_{ij}\frac{8 b^2 N^2}{\pi^6 d}\frac{1}{\bar{z}_i^2} \mathrm{Pe}_m \bar{\xi}^{(1)}_{iim}
\label{eq:rr-2-app-Pe}
\end{align}
The two expressions Eqs.~\eqref{eq:rr1-Pe},~\eqref{eq:rr-2-app-Pe} are equivalent to those in Eqs.~\ref{eq:exp-text}.

\subsection{Polymer length \& stretching coefficient}
\label{sec:expansion_Ls}
In the continuum model the total length of the polymer is not fixed and the polymer can be stretched. Since for real polymers this is rarely the case, we have computed the corrections that should be added to the "stretching" parameter $D$ (see Eq.~\eqref{eq:Rouse}) in order to keep the length of the polymer fixed. In principle this may be important as highlighted in Ref.~\cite{harnau1995dynamic}. 
The average length of the polymer is defined as  $L=\left\langle \sum\limits_{i=1}^{N-1} |\mathbf{r}_{i+1}-\mathbf{r}_i|\right\rangle$, that in the continuum limit reads
\begin{align}
 L=\left\langle \int_{-N/2}^{N/2} \sqrt{\left( \frac{\partial_s \mathbf{r}(t)}{\partial s} \right)^2}ds\right\rangle
\end{align}
\begin{figure}
 \includegraphics[width=0.35\textwidth]{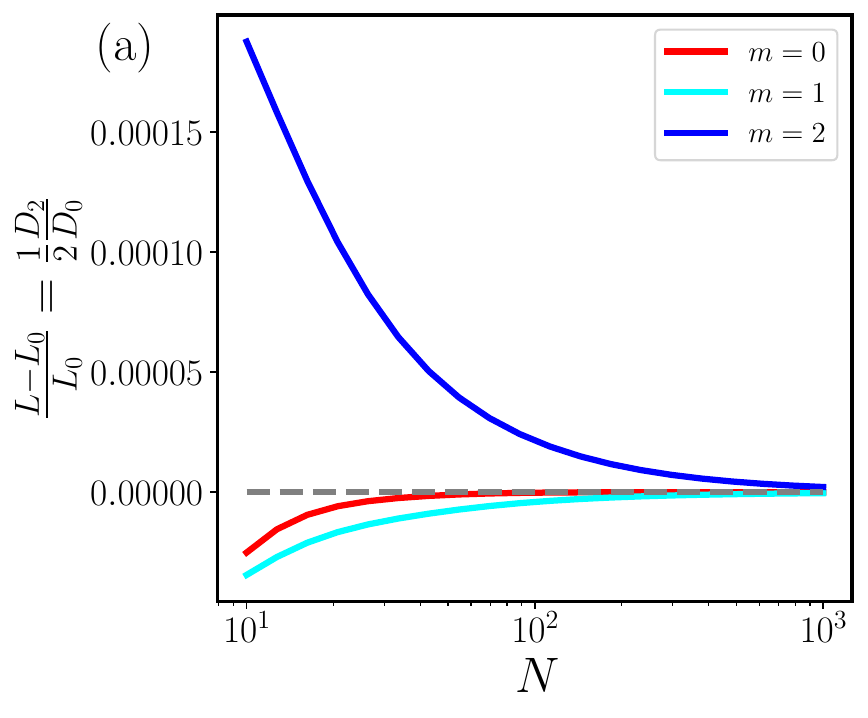}
 \includegraphics[width=0.35\textwidth]{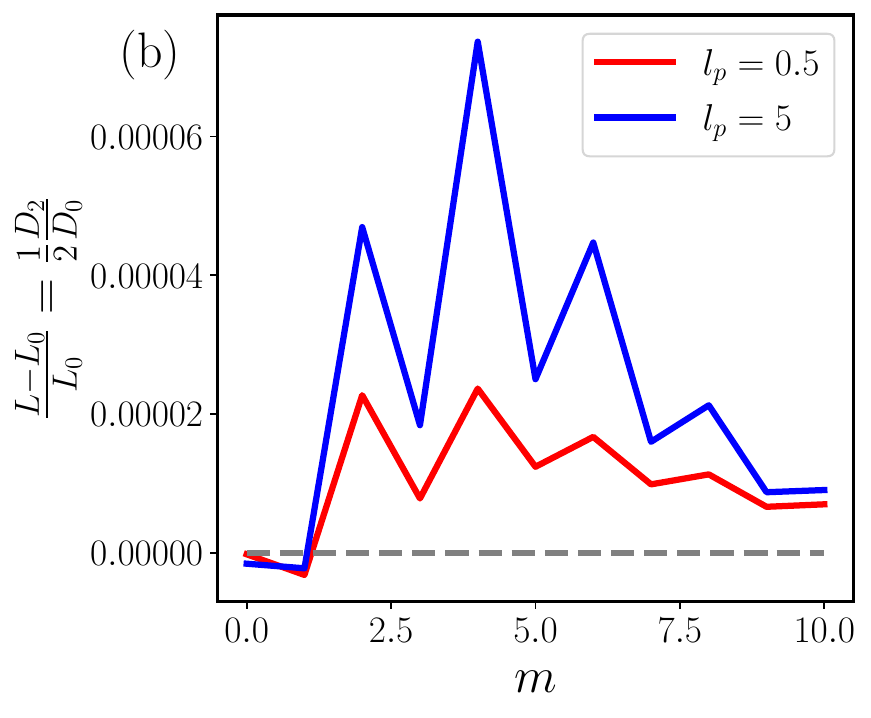}
 \caption{(a): Elongation of the polymer $(L-L_0)/L_0$ (with $L_0=Nb$) as function of $N$ for a single-mode force for $m=0,1,2$ and for $Pe_m = 1$. We recall that due to Eqs.~\eqref{eq:L},\eqref{eq:app-D2} this is also equal to the normalized correction to the stretching coefficient $D_2/D_0$. (b): Elongation of the polymer $(L-L_0)/L_0$ (with $L_0=Nb$) as function of the force mode for $N=100$, $l_p = 0.5,5$ as reported in the legend, and $f_m=1$. }
 \label{fig:L_VS_N}
\end{figure}
{By imposing the constancy of $L$ we can derive the corrections to the effective stretching coefficient that should be accounted for in order to keep $L$ constant 
\begin{align}
    \bar{D} = D_0 + D_1 + D_2
    \label{eq:def-exp_D}
\end{align}
As shown in Appendix \ref{sec:expansion_Ls}, $\bar{D}_1 =0$ and the leading corrections are hence quadratic in the force.} 
We report the ration between the length $L$, computed using Eq.~\eqref{eq:L} over the nominal value $L_0=Nb$, as a function of the number of monomers $N$ for different modes $m$ in Fig.~\ref{fig:L_VS_N}a. We observe that, in the current approximation, the polymer length is barely affected by the activity and it remains almost constant. In particular, $L$ diminishes for $m\leq 1$. Instead, it increases for $m \geq 2$ indicating that the presence of forces with alternating sign will induce a stretching of the chain.

Finally, we provide an analytical expansion of $L(s)$. The average length of the polymer is defined as \\ $L=\left\langle \sum\limits_{i=1}^{N-1} |\mathbf{r}_{i+1}-\mathbf{r}_i|\right\rangle$, that in the continuum limit reads 
\begin{align}
 L=\left\langle \int_{-N/2}^{N/2} \sqrt{\left( \frac{\partial_s \mathbf{r}(t)}{\partial s} \right)^2}ds\right\rangle
\end{align}
Using the eigenfunction representation, 
Eq.~\eqref{eq:theta-exp} and the mean-field approximation we get:
\begin{align}
 L=\int_{-N/2}^{N/2} \sqrt{\langle \Theta(s) \rangle_0}\left(1+\frac{1}{2}\frac{\langle \Theta(s) \rangle_1}{\langle \Theta(s) \rangle_0}+\frac{1}{2}\frac{\langle \Theta(s) \rangle_2}{\langle \Theta(s) \rangle_0}+\frac{1}{4}\left(\frac{\langle \Theta(s) \rangle_1}{\langle \Theta(s) \rangle_0}\right)^2 \right)ds
 \label{eq:L-def1}
\end{align}

In particular,  $\langle \Theta(s) \rangle_0$ (see Eq.~\eqref{eq:Theta0_app-1}) does not depend on $s$, in the limit of small active forces and persistence length.
The expression Eq.~\eqref{eq:L-def1} can be simplified 
\begin{align}
 L=\sqrt{\frac{d}{2}\frac{k_BT}{D l_p}}\left[N+\int_{-N/2}^{N/2} \frac{1}{2}\frac{\langle \Theta(s) \rangle_2}{\langle \Theta(s) \rangle_0}+\frac{1}{4}\left(\frac{\langle \Theta(s) \rangle_1}{\langle \Theta(s) \rangle_0}\right)^2 ds\right]
 \label{eq:L}
\end{align} 
as $\int_{-N/2}^{N/2} \langle \Theta(s) \rangle_1 ds =0$. The other two integrals in Eq.~\eqref{eq:L} can be formally solved
\begin{align}
 \int_{-N/2}^{N/2} \left(\langle \Theta(s) \rangle_1\right)^2  ds&=\frac{4 d}{\pi^4}\frac{k_BT l_p N}{D^3}\mathcal{T}(\{f\},l_p)\\
 \int_{-N/2}^{N/2} \langle \Theta(s) \rangle_2 ds&=\sum_n k_n^2 \langle \mathbf{r}_n \cdot \mathbf{r}_n\rangle^\infty_2=\frac{2}{\pi^4}\frac{l_p N}{D^2}\left[4 N\mathcal{W}(\{f\},l_p)+6l_p \mathcal{H}(\{f\},l_p)\right] \nonumber
 \end{align}
with 
\begin{align}
\mathcal{T}(\{f\},l_p)&=\sum_{m} f_m \sum_{l}\ f_{l} \sum_{i,j\neq i}\sum_{n,q\neq n} \left(\frac{\bar{\xi}_{nqm}^{(0)}}{\bar{z}_{q}}+\frac{\bar{\xi}_{qnm}^{(0)}}{\bar{z}_n}\right)\frac{1}{\bar{z}_n+\bar{z}_{q}}\left(\frac{\bar{\xi}_{ijl}^{(0)}}{\bar{z}_j}+\frac{\bar{\xi}_{jiq}^{(0)}}{\bar{z}_i}\right)\frac{1}{\bar{z}_i+\bar{z}_j} (i\cdot j\cdot n\cdot q)\mathcal{L}_{ijnq}\\
 \mathcal{L}_{ijnq}&=\frac{8}{N}\int_0^N \sin(k_i s) \sin(k_j s) \sin(k_n s) \sin(k_{q} s)ds\\
 \mathcal{W}(\{f\},l_p)&=\sum_n \sum_{m}\sum_{l \leq m} n^2 f_{m} f_{l}   \frac{2 - \delta_{lm}}{2} \left( \sum_{i\neq n}\frac{1}{\bar{z}_i+\bar{z}_n}\left[\frac{\bar{\xi}_{nil}^{(0)}}{\bar{z}_n}+\frac{\bar{\xi}_{inl}^{(0)}}{\bar{z}_i}\right]\bar{\xi}_{nim}^{(0)} +  \sum_{i\neq n}\frac{1}{\bar{z}_i+\bar{z}_n}\left[\frac{\bar{\xi}_{nim}^{(0)}}{\bar{z}_n}+\frac{\bar{\xi}_{inm}^{(0)}}{\bar{z}_i}\right]\bar{\xi}_{nil}^{(0)}  \right) \\
 \mathcal{H}(\{f\},l_p)&=\sum_n \frac{1}{\bar{z}^2_n}\sum_{m} f_m \bar{\xi}^{(1)}_{nnm} 
\end{align}
Notice that these results hold for a generic inhomogeneous active force.
Note also that $\mathcal{L}_{ijnq}\neq 0$ only when $\pm i\pm j\pm q\pm n=0$. It is also interesting to note that, at $f_m=0$ and using the definition of $D$ Eq.~\eqref{eq:def-D}, we retrieve the expected equilibrium value $L=bN$; for $f_m\neq 0$ the polymer becomes stretched or compressed.
However, since real polymers have a fixed contour length, 
we expand the stretching coefficient, $D$, in powers of $f_m$ as $D = D_0 + D_1 + D_2$ 
with 
\begin{align}
    D_0 = \frac{d}{2}\frac{k_BT}{l_p b^2}
\end{align}
and defining $D_1$ and $D_2$ as those values for which $\Delta L=L-bN = 0+\mathcal{O}(f^3)$. 
Accordingly we get 

\begin{align}
 \overline{D}_1&=0\\
 \overline{D}_2&= 2\frac{D_0}{N} \int_0^N \frac{1}{2}\frac{\langle \Theta(s) \rangle_2}{\langle \Theta(s) \rangle_0}+\frac{1}{4}\left(\frac{\langle \Theta(s) \rangle_1}{\langle \Theta(s) \rangle_0}\right)^2 ds\label{eq:app-D2}
\end{align}

\section{Derivation of the expressions for $\mathcal{R}_G$, Eq.~\eqref{eq:def-RG}, and $\mathcal{R}_E$, Eq.~\eqref{eq:def-REE}}\label{app:der-RG}
The gyration tensor is defined as:
\begin{equation}
    \langle S_{\alpha \beta }(t)\rangle=\left\langle \frac{1}{N}\int_{-N/2}^{N/2} (r_{\alpha}(s,t)-r_{0,\alpha}(t))(r_{\beta}(s,t)-r_{0,\beta}(t)) ds\right\rangle
\end{equation}
where $\mathbf{r}_0(t)$ is the location of the center of mass. In the cosines representation $S$ reads:
\begin{multline}
    \langle S_{\alpha \beta}(t)\rangle=\left\langle\frac{2}{N^2}\!\int_{-N/2}^{N/2} \sum_{i,j} r_{i,\alpha}(t)\,r_{j,\beta}(t)\,\cos(k_{i} s)\cos(k_{j} s) - \sum_{i}  \left[r_{0,\alpha}(t)\,r_{i,\beta}(t)+r_{0,\beta}(t)\,r_{i,\alpha}(t)\right]\cos(k_{i} s)  +r_{0,\alpha}\,r_{0,\beta}\,ds \right\rangle
\end{multline}
that finally  reads:
\begin{equation}
    \langle S_{\alpha \beta}(t)\rangle=\frac{1}{N}\left[\sum_{i}  \langle r_{i,\alpha}(t)\,r_{i,\beta}(t)\rangle-\langle r_{0,\alpha}\,r_{0,\beta}\rangle\right]=\frac{1}{N}\left[\sum_{i\neq 0}  \langle r_{i,\alpha}(t)\,r_{i,\beta}(t)\rangle\right]
    \label{eq:gyr-ten}
\end{equation}
The gyration radius is the trace of $S$, and at steady state it reads
\begin{align}
\mathcal{R}^2_G=\text{Tr}S^\infty=\frac{1}{N}\sum_{i\neq 0}\sum_{\alpha}\langle r_{i,\alpha}\,r_{i,\alpha}\rangle^\infty = \frac{1}{N}\sum_{i\neq 0}\langle \mathbf{r}_{i}\cdot \mathbf{r}_{i}\rangle^\infty
\end{align}

The end-to-end square distance is defined as 
\begin{align}
\mathcal{R}^2_E &=\left\langle (\mathbf{r}_{-N/2}-\mathbf{r}_{N/2})\cdot (\mathbf{r}_{-N/2}-\mathbf{r}_{N/2})\right\rangle  
\end{align}
which in the continuum representation reads
\begin{align}
\mathcal{R}^2_E &= \int_{-N/2}^{N/2} ds\int_{-N/2}^{N/2} dw\langle \mathbf{r}'(s)\cdot \mathbf{r}'(w)\rangle^\infty
\end{align}
where $\mathbf{r}'(s)$ is the derivative of $\mathbf{r}(s)$ with respect to the position on the backbone, $s$. In the small force and small persistence length approximation, the last expression reads
\begin{align}
\mathcal{R}^2_E &= \frac{2}{N}\int_{-N/2}^{N/2} ds\int_{-N/2}^{N/2} dw\sum_{i=0}^{\infty}\sum_{j=0}^{\infty}\langle  \mathbf{r}_{i} k_{i} \sin(k_{i} s) \cdot \mathbf{r}_{j} k_{j} \sin(k_{j} w)  \rangle^\infty\nonumber\\
 & =  \frac{8}{N}\sum_{i=odd}\sum_{j=odd} \langle \mathbf{r}_{i}\cdot\mathbf{r}_{j}\rangle^\infty
 \label{app:def-REE}
\end{align}

\section{Numerical results on negative active forces}
\label{sec:appendix_neg}

\begin{figure}[!h]
 \includegraphics[width=0.7\textwidth]{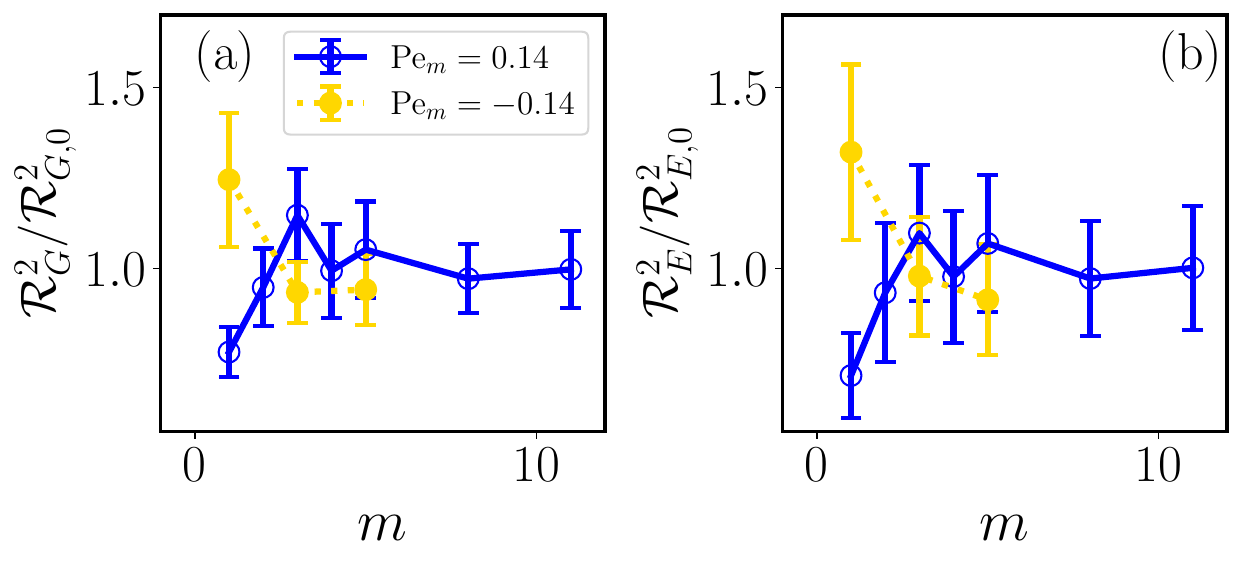}
 \caption{(a) Ratio of the square gyration radius $\mathcal{R}^2_{G}$, over the equilibrium value $\mathcal{R}^2_{G,0}$ and (b) ratio of the end-to-end square distance $\mathcal{R}^2_{E}$, over the equilibrium value $\mathcal{R}^2_{E,0}$ as a function of the mode $m$ and different values of $\mathrm{Pe}_m$ (fixed absolute value, one positive and one negative) at fixed $N=200$ from numerical simulations. }
 \label{fig:check_neg}
\end{figure}

We report the comparison between numerical simulations performed at small values of $\mathrm{Pe}_m$, where we compare the effect of flipping the sign of the force while keeping its absolute value fixed. Results are shown in Fig.~\ref{fig:check_neg}, where we report $\mathcal{R}^2_{G}/\mathcal{R}^2_{G,0}$ and $\mathcal{R}^2_{E}/\mathcal{R}^2_{E,0}$ and focus on odd force modes $m=1,3,5$. As suggested in the main text, the odd force mode are the only ones that are susceptible to the force flip and, in particular, the mode $m=1$ is the most susceptible, as a sign change from positive to negative would amount to switching from an overall compression to an overall extension. Results for $\mathrm{Pe}_m <0$ at both $m=3$ and $m=5$ show a slight decrease with respect to the corresponding data for $\mathrm{Pe}_m >0$; however, the results are compatible, being within one standard deviation. Instead, for $m=1$ results are not only statistically incompatible but we observe a marked stretching of the polymer for $\mathrm{Pe}_m <0$, thus bringing a qualitative difference with respect to the compression reported for $\mathrm{Pe}_m >0$. As mentioned in the main text, the analytical model, derived for weak active forces, does not predict this discrepancy in the case of the gyration radius.    

\end{document}